\documentclass[12pt]{article}

\usepackage{amssymb}

\title{Deformation Quantisation of Constrained Systems}
\author{\large Frank Antonsen \\University of Copenhagen \\Niels Bohr 
Institute}

\begin{document}
\maketitle

\begin{abstract}
We study the deformation quantisation (Moyal quantisation) of general
constrained Hamiltonian systems. It is shown how
second class constraints can be
turned into first class quantum constraints. This is illustrated by the 
$O(N)$ non-linear $\sigma$-model. Some new light is also
shed on the Dirac bracket. Furthermore, it is shown how classical constraints
not in involution with the classical Hamiltonian, can be turned into quantum
constraints {\em in} involution with respect to the Hamiltonian. 
Conditions on the existence of anomalies
are also derived, and it is shown how some kinds of anomalies can be removed.\\
The equations defining the set of physical states are also given.\\
It turns out that the deformation quantisation of pure Yang-Mills theory
is straightforward whereas gravity is anomalous. A formal solution to the
Yang-Mills quantum constraints is found. In the 
\small{ADM} formalism of gravity the anomaly is very complicated and the 
equations 
picking out physical states become infinite order functional differential
equations, whereas the Ashtekar variables remedy both of these problems --
the anomaly becoming simply a central extension (Schwinger term) and the
equations for physical states become finite order.\\
We finally elaborate on the underlying geometrical structure and show the
method to be compatible with BRST methods.
\end{abstract}

\section{Introduction}
The problem of how to quantise a given classical system is one of the oldest
in the history of quantum theory, its age being comparable to that of the
measurement problem. This being so, many attempts have been made in the
past of defining general quantisation schemes. Usually such schemes try
to construct mappings $q_i,p^j\mapsto \hat{q}_i,\hat{p}^j$ such that
\begin{equation}
	\{ f(q,p), g(q,p) \}_{\rm PB} \mapsto \frac{1}{i\hbar} 
	[\hat{f}(\hat{q},\hat{p}),\hat{g}(\hat{q},\hat{p})]
\end{equation}
where $\{\cdot,\cdot\}_{\rm PB}$ denotes the Poisson bracket,
\begin{equation}
	\{ f(q,p), g(q,p) \}_{\rm PB} = \frac{\partial f}{\partial q_i}
	\frac{\partial g}{\partial p^i} - \frac{\partial f}{\partial p^i}
	\frac{\partial g}{\partial q_i} 
\end{equation}
and where $\hat{f}$ denotes some kind of operator-valued function obtained
from the classical observable $f$ in some prescribed manner. The most elaborate
of these schemes must be geometric quantisation, \cite{geomq}, which relies
heavily on the symplectic geometry of classical phasespaces.\\
Unfortunately, one can show that in general no quantisation procedure can
exists for functions which are not at most quadratic polynomials in the
basic variable, $q_i,p^j$, \cite{geomq}. This implies that the algebra will
receive quantum corrections. In fortunate circumstances these are just central
extensions. \footnote{One should note that BRST does not escape this problem
either - given a classical BRST generator, the task is yet again to find
an operator realisation. The problem is lightened a little bit, however, since
BRST exploits the structure of the classical theory to a far greater extent 
than old fashioned canonical quantisation. Ultimately, though, it ends up with
similar problems.\\
Geometric quatisation avoids the no-go theorem by loosening another of Dirac's
requirements, namely that $q,p$ be irreducibly represented. This can lead to
problems with the classical limit, however.}\\
Another procedure, which has not been studied as much as geometric 
quantisation, is {\em deformation quantisation}. It is known that this always
works, \cite{deform} (recently, Kontsevich has extended the proof of 
existence from symplectic to Poisson manifolds too, \cite{Kontsevich}). 
Denote the classical phasespace by $\Gamma$, this is
then a symplectic manifold, and the set of observables, $C^\infty(\Gamma)$,
form a Poisson-Lie algebra (i.e., a Lie algebra under Poisson brackets), 
$\cal A$.
Deformation quantisation consists in replacing the algebra of observables
with another, deformed one, ${\cal A}_\hbar$, and the Poisson brackets 
by a new, deformed bracket. More concretely, ${\cal A}_\hbar 
= {\cal A}\otimes \mathbb{C}((\hbar))$, i.e., the elements of ${\cal A}_\hbar$
are functions on
phasespace which can also be expanded in a power series in $\hbar, 
\hbar^{-1}$.\footnote{Standard mathematical notation: $\mathbb{C}[x]$ 
denotes the
set of complex polynomials in one variable $x$, $\mathbb{C}[[x]]$ the set of 
formal
power series in $x$, $\mathbb{C}(x)$, the field of fractions of 
$\mathbb{C}[x]$ (i.e., $\mathbb{C}(x)
=\{f(x)/g(x)~|~ f,g\in \mathbb{C}[x], g(x)\neq 0\}$, and $\mathbb{C}((x))$ 
the field of
fractions of $\mathbb{C}[[x]]$, i.e., $\mathbb{C}((x)) \simeq 
\mathbb{C}[[x,x^{-1}]]$.} Thus elements of ${\cal A}_\hbar$ can be written as
\begin{equation}
	f(q,p,\hbar) = ... + \hbar^{-1} f_{-1}(q,p) + f_0(q,p) + \hbar f_1(q,p)
	+...
\end{equation}
where $f_n\in {\cal A}=C^\infty(\Gamma), n\in\mathbb{Z}$. Here $f_0$ is the 
{\em classical part} of $f$. Similarly one writes the new bracket as
\begin{equation}
	[f,g]_M = i\hbar \{f,g\}_{\rm PB} + O(\hbar^2)
\end{equation}
We will in general refer to the deformed brackets as {\em Moyal brackets},
even though that name is strictly speaking reserved for deformed brackets 
of flat phasespaces, $\Gamma\simeq \mathbb{R}^{2n}$. It has recently been 
proven by 
Tzanakis and Dimakis, \cite{Tzan}, that the Moyal bracket is essentially
unique -- the various possibilities corresponding to various operator ordering
prescriptions. Consequently, deformation quantisation does not suffer from the 
usual amibiguities of other quantisation schemes. Furthermore, Dereli and
Ver\c{c}in have proven that the various possibilities for the Moyal bracket
du to different operator orderings, satisfy a $W_\infty$-covariance, 
\cite{DerVer}. The close relationship between $W$-symmetry and the 
Wigner-Weyl-Moyal formalism has also been extensively studied by Gozzi and
coworkers, \cite{Gozzi}.\\
Furthermore, we will usually restrict ourselves to the simpler case of only
non-negative powers of $\hbar$, i.e., work with ${\cal A}_\hbar= {\cal A}
\otimes\mathbb{C}[[\hbar]]$. Only when this turns out to be impossible 
will we deal
with the more general case of ${\cal A}\otimes\mathbb{C}((\hbar))$.
Systems with second class constraints or anomalies will in general require
the presence of negative powers of $\hbar$, whereas systems without such 
problems in general will not.\\
The general philosophy of the paper will be that for any classical theory
with problems (such as second class constraints, Hamiltonian not in
involution with the constraints, anomalies appearing upon a simple 
quantisation), a completely well-behaved quantum theory exists for which all
of these problems are absent. The problems only reappear when taking the
classical limit, a procedure which has to be slightly modified. In other
words, we have a ``quantum smoothening out'' of classical problems, analogous
to what happens in quantum chaos, where a chaotic classical Hamiltonian still
leads to an integrable system at the quantum level, because the phasespace
gets ``smurred'' by the Heisenberg uncertainty principle. I think this is a
sensible procedure, as the quantum description is presumably the most
fundamental one, and the classical description should at most be considered
as a limiting case -- and as such, we would expect problems to occur, not
so much at the fundamental level (nature tends to favour simplicity) but 
arrising from taking the limit.\\ 
For $\Gamma\simeq \mathbb{R}^{2n}$ a compact expression for the 
Moyal bracket exists.
Introduce the bidifferential operator $\triangle$,\footnote{We will only in 
this paper deal with bosonic degrees of freedom. Fermionic variables can be 
treated by means of a $\mathbb{Z}_2$-grading, where the usual 
Poisson bracket is 
replaced by a graded analogue, as in the Batalin-Fradkin-Vilkovisky approach 
to \small{BRST}-quantisation.}
\begin{equation}
	f\triangle g = \{f,g\}_{\rm PB}
\end{equation}
then
\begin{equation}
	[f,g]_M = 2if\sin (\frac{1}{2}\hbar\triangle) g
\end{equation}
see e.g. \cite{deform}. This can be written out more explicitly as
\begin{equation}
	[f,g]_M = i \sum_{n=0}^\infty \frac{(-1)^n\hbar^{2n+1}}{4^n(2n+1)!}
	\sum_{k=0}^n (-1)^k\left(\begin{array}{c}n\\k\end{array}\right)
	\frac{\partial^{2n+1} f}{\partial q^{2n-k+1}\partial p^k}
	\frac{\partial^{2n+1} g}{\partial p^{2n-k+1}\partial q^k}
	 \label{eq:Moyal}
\end{equation}
with
\begin{displaymath}
	\left(\begin{array}{c}n\\k\end{array}\right) =\frac{n!}{k!(n-k)!}
\end{displaymath}
denoting a binomial coefficient.\\
The Moyal bracket can be written in terms of a so-called twisted product
\begin{equation}
	f*g = f e^{\frac{1}{2}i\hbar\triangle} g = fg +O(\hbar)
\end{equation}
the Moyal bracket is then the commutator with respect to this product
\begin{equation}
	[f,g]_M = f*g-g*f
\end{equation}
which brings out the analogy with the standard operator formulation of
quantum theory.\\
In general we will write
\begin{equation}
	[f,g]_M = \sum_{n=1}i\hbar^n\omega_n(f,g)
\end{equation}
with $\omega_1(f,g) = \{f,g\}_{\rm PB}$.\\
Since one can always choose local Darboux coordinates, $q_i,p^j$ (with the
standard Poisson bracket) we can always at least locally make do with the
Moyal bracket, without having to work with a more general deformation.
Furthermore, for most systems of physical interest, such as {\em all} 
standard field
theories, we can take the classical phasespace to be ``flat'' in the sense
that it can be covered by a global coordinate patch of Darboux coordinates.
The possible ``curvature'' of $\Gamma$ is put into a set of constraints.
Hence one may extend the original phasespace, which might not have a global
Darboux coordinate patch, into a ``flat'' phasespace $\Gamma'$, such that
$\Gamma$ is characterised by the vanishing of certain functions $\phi_a(q,p)$,
i.e.,
\begin{equation}
	\Gamma = \{(q,p)\in\Gamma' ~|~ \forall a: \phi_a(q,p)=0\}
\end{equation}
This leads us naturally to study Hamiltonian systems with constraints. 

\section{Hamiltonian Systems with Constraints}
Consider a classical Hamiltonian system with a phasespace $\Gamma$ and a
Hamiltonian $h(q,p)$. Suppose the transition from a Hamiltonian to a 
Lagrangian picture is singular, i.e., suppose not all the phasespace 
coordinates are independent. We then have a system with constraints, i.e.,
a family of functions $\phi_a(q,p)$ exists such that the true degrees of
freedom are characterised by the vanishing of these functions. We will also
assume these functions to be independent.\\
We will, at first, assume that the set of constraints are {\em first class}, 
i.e., they satisfy a Poisson-Lie algebra
\begin{equation}
	\{\phi_a,\phi_b\}_{\rm PB} = c_{ab}^{~~c}\phi_c
\end{equation}
where the structure coefficients, $c_{ab}^{~~c}$ can be functions of the
phasespace variables.\\
For now we will also assume that time evolution does not give rise to new 
constraints, i.e., the functions $\phi_a$ are {\em in involution} with the
Hamiltonian
\begin{equation}
	\{h,\phi_a\}_{\rm PB} = V_a^b\phi_b
\end{equation}
where again the structure coefficient $V_a^b$ can depend on the phasespace
coordinates. We will return to the more general setting in a subsection 
later.\\
Now we want to perform a deformation quantisation. We will thus replace
$\phi_a$ by functions $\Phi_a=\phi_a+O(\hbar,\hbar^{-1})$, the Hamiltonian
$h$ with $H=h+O(\hbar,\hbar^{-1})$ and the Poisson brackets with Moyal ones,
in such a manner that
\begin{eqnarray}
	[\Phi_a,\Phi_b]_M &=& i\hbar c_{ab}^c\Phi_c \label{eq:Ma1}\\
	\left[H,\Phi_a\right]_M &=& i\hbar V_a^b\Phi_b \label{eq:Ma2}
\end{eqnarray}
For simplicity we will only consider $\Phi_a,H$ to be formal powerseries in
$\hbar$ and not $\hbar^{-1}$. Write
\begin{equation}
	\Phi_a = \phi_a+\sum_{n=1}^\infty \hbar^n \Phi_a^{(n)}\qquad
	H = h+\sum_{n=1}^\infty\hbar^nH_n
\end{equation}
Furthermore we have already assumed that the structure coefficients are
undeformed. When the Moyal bracket algebra, (\ref{eq:Ma1}-\ref{eq:Ma2}),
become non-isomorphic to the
classical Poisson bracket one we say we have an {\em anomaly}. This can happen
if e.g. the structure constants get $\hbar$-corrections, the algebra become
centrally extended or if the right hand sides of (\ref{eq:Ma1}-\ref{eq:Ma2})
become non-linear in the $\Phi_a$. Such problems will be dealt with in a
later section.\\
At each order in $\hbar$ the Moyal algebra relations (\ref{eq:Ma1}-
\ref{eq:Ma2}) read
\begin{eqnarray}
	\sum_{n,k}\omega_{N-n-k}(\Phi_a^{(n)},\Phi_b^{(k)}) &=& c_{ab}^c
	\Phi_c^{(N)}\\
	\sum_{n,k}\omega_{N-n-k}(H_n,\Phi_a^{(k)}) &=& V_a^b\Phi_b^{(N)}
\end{eqnarray}
which provides us with a recursive scheme for finding the quantum constraints
$\Phi_a$ and the quantum Hamiltonian $H$. Inserting the explicit form for
$\omega_l$ we get
\begin{eqnarray}
	\sum_{n,k} \sum_{l=0}^{N-n-k} \frac{(-1)^{N-n-k+l}}{4^{N-n-k} 
	(2N-2n-2k+1)!} \left(\begin{array}{c}N-n-k\\l\end{array}\right)
	\times&&\nonumber\\
	\frac{\partial^{N-n-k}\Phi_a^{(n)}}{\partial p^{N-n-k-l}\partial q^l}
	\frac{\partial^{N-n-k}\Phi_b^{(k)}}{\partial q^{N-n-k-l}\partial p^l}
	&=& c_{ab}^c\Phi_c^{(N)}\\
	\sum_{n,k} \sum_{l=0}^{N-n-k} \frac{(-1)^{N-n-k+l}}{4^{N-n-k} 
	(2N-2n-2k+1)!} \left(\begin{array}{c}N-n-k\\l\end{array}\right)
	\times&&\nonumber\\
	\frac{\partial^{N-n-k}H_n}{\partial p^{N-n-k-l}\partial q^l}
	\frac{\partial^{N-n-k}\Phi_a^{(k)}}{\partial q^{N-n-k-l}\partial p^l}
	&=& V_a^b\Phi_b^{(N)}
\end{eqnarray}
One should note that if $\phi_a,h$ are all at most third order in $q,p$ then
$\omega_n, n\geq 3$ on these vanish identically and one can take 
simply $\Phi_a=\phi_a,
H=h$. Many physical systems have precisely this structure. Most notably
the Yang-Mills field theory. The study of membranes or some supersymmetrical
theories, however, will in general need $\hbar$-corrections to the constraints
and Hamiltonian.\\
It is well worth having a closer look at the recursive scheme above. The first
equation, the one for the first quantum correction, reads
\begin{equation}
	c_{ab}^c\Phi_c^{(1)} = \{\phi_a,\Phi_b^{(1)}\}_{\rm PB} +
	\{\Phi_a^{(1)},\phi_b\}_{\rm PB}
\end{equation}
which gives a first order partial differential equation for $\Phi_a^{(1)}$,
\begin{equation}
	\left[c_{ab}^c+\delta_a^c\left(\frac{\partial\phi_b}{\partial q}
	\frac{\partial}{\partial p}-\frac{\partial\phi_b}{\partial p}
	\frac{\partial}{\partial q}\right) 
	-\delta_b^c\left(\frac{\partial\phi_a}{\partial q}
	\frac{\partial}{\partial p}-\frac{\partial\phi_a}{\partial p}
	\frac{\partial}{\partial q}\right)\right]\Phi_c^{(1)} =0
\end{equation}
Introduce the operator
\begin{equation}
	{\cal D}_{ab}^c = c_{ab}^c+\delta_a^c\{\phi_b,\cdot\}_{\rm PB}
	-\delta_b^c\{\phi_a,\cdot\}_{\rm PB}
\end{equation}
we can then write the equation for $\Phi_c^{(1)}$ as
\begin{equation}
	{\cal D}_{ab}^c\Phi_c^{(1)} = 0 \label{eq:calD}
\end{equation}
The other equations in the hierarchy have a similar form, namely
\begin{equation}
	{\cal D}_{ab}^c\Phi_c^{(n)} = {\cal J}_{ab}^{(n)} \label{eq:hier}
\end{equation}
where ${\cal J}_{ab}^{(n)}$ is a source term only depending on $\Phi_c^{(0)}=
\phi_c,\Phi_c^{(1)},...,\Phi_c^{(n-1)}$. Explicitly
\begin{equation}
	{\cal J}_{ab}^{(n)} = -\sum_{m=1}^n\sum_{k=1}^{m+1}\omega_m
	(\Phi_a^{(k)},\Phi_b^{(n+1-m-k)})
\end{equation}
which shows how the ``source'' is build up from the various $m$-order
brackets, $\omega_m$.\\
Note, by the way, that if (\ref{eq:calD}) has only the trivial solution
$\Phi_c^{(1)}\equiv 0$, then it follows from (\ref{eq:hier}) that $\Phi_c^{(n)}
\equiv 0, n\geq 1$ and hence $\Phi_a=\phi_a$. This would of course not be the
case if we included negative powers of $\hbar$ in the expansion of $\Phi_c$.
In any case equations (\ref{eq:calD}-\ref{eq:hier}) show that $\Phi_c^{(n)}$
cannot be uniquely defined as we can always add a multiple of $\Phi_c^{(1)}$
to it (with an appropriate factor of $\hbar$ in front, naturally).\\
Furthermore, the operator ${\cal D}_{ab}^c$ cannot be identically zero,
since the classical constraints satisfy
\begin{equation}
	{\cal D}_{ab}^c\phi_c = -c_{ab}^{~~~c}\phi_c
\end{equation}
and are hence ``eigenvectors'' of ${\cal D}_{ab}^c$.
Equation (25) constitue a set of first order partial differential equations
determining the quantities $\Phi_a^{(n)}$.\\
There is a cohomological element to this set of equations. Denote the
algebra of constraints by $\mathfrak{g}$, $\mathfrak{g}\subseteq {\cal A}=C^\infty(\Gamma)$,
and form the complex ${\cal C}^n = \wedge^n\mathfrak{g}$ of alternating $n$-linear 
expressions in the
constraints $\phi_a$. Introduce the operator $\hat{A}_{ab}^{~~~c}: {\cal C}^1
\rightarrow{\cal C}^2$ by
\begin{equation}
	\hat{A}_{ab}^{~~c} f_c := \{\phi_a,f_b\}_{\rm PB} - \{\phi_b,f_a\}_{\rm
	PB}
\end{equation}
We extend this by linearity to a map ${\cal C}^p\rightarrow{\cal C}^{p+1}$,
\begin{equation}
	\hat{A}_{ab}^{~~c_1} f_{c_1...c_p} := \sum_{\sigma\mbox{ cyclic}}
	{\rm sign}(\sigma)\{\phi_{\sigma(a)},f_{\sigma(b)\sigma(c_2)...
	\sigma(c_p)}\}_{\rm PB}
\end{equation}
where the $\sigma$ is a cylic permutation of the letters $a,b,c_2,...,c_p$, 
i.e., $\sigma\in S_{p+1}$. Then by construction
\begin{eqnarray}
	\hat{A} &:& {\cal C}^p\rightarrow {\cal C}^{p+1}\\
	\hat{A}^2 &=& 0
\end{eqnarray}
Hence we have a cohomology theory
\begin{eqnarray}
	\hat{Z}^p &:=& \left.{\rm Ker}\hat{A}\right|_{{\cal C}^p}\\
	\hat{B}^p &:=& \left.{\rm Im}\hat{A}\right|_{{\cal C}^{p-1}} = \hat{A}
	{\cal C}^{p-1}\\
	\hat{H}^p &:=& \hat{Z}^p/\hat{B}^p
\end{eqnarray}
By construction $\hat{H}^1$ measures the obstruction to invertibility of
$\hat{A}$ as an operator on functions, and hence to the uniqueness of 
$\Phi_c^{(1)}$. The cohomology cannot be entirely trivial since
\begin{equation}
	\hat{A}_{ab}^{~~c}\phi_c = 2c_{ab}^{~~c}\phi_c
\end{equation}
i.e., the constraints themselves are a kind of eigenvector system for 
$\hat{A}$.\\
Now, the standard complex computing the cohomology of a Lie algebra is the
Koszul complex which is precisely our ${\cal C}^p$ complex, 
\cite{Jacobson}. The derivative operator, however, is given by
\begin{equation}
	sf(x_1,...,x_{p+1}) = \sum_{q<r}(-1)^{q+r}f(x_1,...,\hat{x_q},...
	,\hat{x}_r,...,x_{p+1},[x_q,x_r])
\end{equation}
where a hat over an $x$ denotes that it is to be omitted and where $[\cdot,
\cdot]$ denotes the appropriate Lie bracket (here the Poisson bracket). But
the Poisson bracket is also a derivation, i.e., satisfies the
Leibnitz rule
\begin{equation}
	\{x,yz\}_{PB} = \{x,y\}_{\rm PB}z+y\{x,z\}_{\rm PB}
\end{equation}
This is in fact one of the defining relations of a Poisson algebra,
namely that it is a Lie algebra whose Lie bracket also satisfies the Leibnitz
rule. From this one immediatly sees that $\hat{A}=s$ (possibly up to a sign),
since
\begin{eqnarray*}
	\{\phi,f^{1...p}x_1...x_p\}_{\rm PB} &=& f^{1...p}\{\phi,x_1...x_p
	\}_{\rm PB}\\
	&=& f^{1...p}\sum_q\{\phi,x_q\}_{\rm PB}x_1...\hat{x}_q...x_p
\end{eqnarray*}
Consequently $\hat{H}^*$ is nothing but the usual Lie algebra cohomology.\\
In a somewhat compact, symbolic notation the condition (\ref{eq:cond}) then
implies that $c\Phi$ and ${\cal J}$ differ by an exact term. Explicitly
\begin{equation}
	c_{ab}^c\Phi^{(1)}_c = \{\phi_a,\phi_b\}_{\rm PB} +\mbox{exact form} 
\end{equation}
etc. For Lie algebras with vanishing cohomology we then see that the higher
order terms $\Phi^{(n)}_a$ are completely determined by the $\omega_k$
brackets on the lower order terms. This ensures existence of such quantum
constraints. It furthermore shows that equivalent quantum constraints only 
differ by exact terms, i.e., two set of quantum constraints are equivalent
(give the same Moyal algebra) if and only if they are cohomologous (i.e., that
they at each power of $\hbar$ only differ by an exact form).\\
A closely related cohomology generator will appear again when we deal
with second class constraints.\\
This procedure of simply replacing Poisson brackets by Moyal ones has been 
used by Pleba\'{n}ski and coworkers in the study of self-dual gravity, large
$N$ limit of $SU(N)$ theories and Nahm equations, \cite{Plebanski}. The only
new thing here is the level of generality -- instead of considering a 
particular theory with constraints, we here deal with a generic situation.

\subsection{Some Particular Solutions}
It is quite natural to assume $\Phi_a\approx\phi_a$ (where $\approx$ denotes
weak equivalence in the sense of Dirac), or more concretely that $\Phi_a$
only depends on $(q,p)$ through the classical constraints $\phi_a$. Assuming
this, the first equation in the recursive scheme reads
\begin{equation}
	c_{ab}^c\Phi_c^{(1)}(\phi) = c_{ac}^d\phi_d\frac{\partial\Phi_b^{(1)}}
	{\partial\phi_c}-c_{bc}^d\phi_d\frac{\partial\Phi_a^{(1)}}{\partial
	\phi_c}
\end{equation}
a solution of which is $\Phi_c^{(1)} = \alpha_c^{ab}\phi_a\phi_b$ with 
$\alpha_c^{ab}$ a constant satisfying
\begin{eqnarray}
	\alpha_c^{ab} &=& \alpha_c^{ba}\\
	2\alpha_b^{ef}c_{ae}^g-2\alpha_a^{ef}c_{be}^g &=& 
	c_{ab}^c\alpha_c^{gf}
\end{eqnarray}
This set of equations cannot always be guaranteed to have a non-trivial
solution. Abelian constraint algebras are obviously not a problem, so let
us consider the simplest non-Abelian algebra
\begin{displaymath}
	\{\phi_1,\phi_2\}_{\rm PB} = \phi_2
\end{displaymath}
in this case the structure coefficient is $c_{12}^2=1$ and the equation for
$\alpha_c^{ab}$ reduces to
\begin{displaymath}
	\left(\delta_{a1}\delta_{b2}-\delta_{a2}\delta_{b1}\right)
	\alpha_2^{gf}
	=2\left(\alpha_b^{2f}\delta_{a1}-\alpha_b^{1f}\delta_{a2}-
	\alpha_a^{2f}\delta_{b1}+\alpha_a^{1f}\delta_{b2}\right)
	\delta^{g2}
\end{displaymath}
which has the solution
\begin{eqnarray*}
	\alpha_f^{22} &=& \mbox{ arbitrary}\\
	\alpha_2^{1f} &=& \alpha_1^{11} =0\\	
	\alpha_1^{12} &=& -\frac{1}{2}\alpha_2^{22}
\end{eqnarray*}
whence the quantum constraints read	
\begin{eqnarray*}
	\Phi_1 &=& \phi_1 +\hbar\left(-\frac{1}{2}\phi_1+\alpha_1^{22}
	\phi_2\right)\phi_2\\
	\Phi_2 &=& \phi_2 +\hbar\alpha_2^{22}\phi_2^2
\end{eqnarray*}
This then constitute an example with $\Phi_a\approx \phi_a$, where the 
quantum constraints depend on the phasespace variable only through the
classical constraints. One should note that such a construction has found
use in general relativity, where a new Abelian constraint replacing the 
Hamiltonian constraint has been found, \cite{Kuchar}. In that case, however,
since the formal manipulations were carried out at a classical level, Planck's
constant did not appear as a deformation parameter. Nevertheless, it is
interesting to note the similarity. We will return to this new gravitational
constraint algebra in a later section.\\
For the case of an $su_2$ algebra, $\{\phi_a,\phi_b\}_{\rm PB} = 2i
\epsilon_{ab}^{~~c}\phi_c$, the conditions on the coefficients $\alpha_a^{bc}$
read
\begin{eqnarray*}
	\alpha_2^{2f}\delta^g_3-\alpha_2^{3f}\delta^g_2+\alpha_1^{1f}\delta^g_3
	-\alpha_1^{3f}\delta^g_1 &=& \frac{1}{2}i\alpha_3^{gf}\\
	\alpha_3^{2f}\delta^g_3-\alpha_3^{3f}\delta^g_2-\alpha_1^{1f}\delta^g_2
	+\alpha_1^{2f}\delta^g_1 &=& -\frac{1}{2}i\alpha_2^{gf}\\
	-\alpha_3^{1f}\delta^g_3+\alpha_3^{3f}\delta^g_3-\alpha_2^{1f}
	\delta^g_2+\alpha_2^{3f}\delta^g_1&=& i\alpha_1^{gf}
\end{eqnarray*}
which only has the trivial solution $\alpha_a^{bc}\equiv 0$, illustrating that
not all algebras can be treated in this way. The example of an $O(N)$ 
non-linear $\sigma$-model belongs to this category as will be shown later.

\subsection{The Physical States}
Now, this treatment of constrained systems have dealt with merely the
kinematics and not the dynamics proper, i.e., we have not studied the
equations of motion. In the standard Dirac treatment of constrained Hamiltonian
systems the first class constraints become operators upon quantisation, i.e.,
physical states have to satisfy
\begin{equation}
	\hat{\phi}_a|\Psi\rangle =0 \qquad \forall a
\end{equation}
where $\hat{\phi}_a$ is some operator realisation of $\phi_a$. How is this
modified in a deformation quantisation?\\
A good starting point is to consider the deformation quantisation analogue
of a state $|\Psi\rangle$, namely a {\em Wigner function} $W_\Psi$. 
We could then impose the constraints in the following algebraic way:
\begin{displaymath}
	\Phi_a*W_\Psi = 0\qquad \forall a
\end{displaymath}
or more symmetrically (where $[\cdot,\cdot]_M^+$ denotes the ``anti-Moyal
bracket'', $[f,g]_M^+ := f*g+g*f = 2 f\cos(\frac{1}{2}\hbar\triangle)g$)
\begin{equation}
	[\Phi_a,W_\Psi]_M^+ = 0 \qquad\forall a \label{eq:cond}
\end{equation}
Notice,
\begin{equation}
	0=[\Phi_a,W_\Psi]_M^+ = 2\Phi_a W_\Psi+...
\end{equation}
so to lowest order we have $W_\Psi\equiv 0$ away from the constraint
surface. More precisely
\begin{equation}
	{\rm supp} W_\Psi^{(0)} \subseteq \bigcap {\rm Ker} \phi_a
\end{equation}
where $W_\Psi^{(0)}$ is the $\hbar^0$ component of $W_\Psi$.\\
The physical Hilbert space, ${\cal H}_{\rm phys} \equiv\{|\Psi\rangle ~|~
\forall a:\hat{\phi}_a|\Psi\rangle=0\}$, then gets replaced by the space
\begin{equation}
	{\cal C}_{\rm phys} := \{W~|~\forall a: [\Phi_a,W]_M^+=0\}
\end{equation}
This also has the advantage of
being analogous to the \small{BRST}-condition, $[\hat{\Omega},A]=0$ where
$[\cdot,\cdot]$ is a graded commutator, $\hat{\Omega}$ is the 
\small{BRST}-charge, $\hat{\Omega} = \eta^a\hat{\phi_a} +...$ and $A$ is any
observable (the observable corresponding to a state $|\Psi\rangle$ is
of course the projection operator $A=|\Psi\rangle\langle\Psi|$). Thus we will
be using (\ref{eq:cond}) as defining the deformation quantisation analogue of
the Dirac condition.\\
A few comments are in order. First, the replacement of Poisson brackets by
Moyal ones implies that the corresponding ``gauge'' transformations acquire
quantum modifications. If the classical constraints are $\phi_a$, then they
generate (infinitesimal) ``gauge'' transformations $\delta_\omega f:= 
\{\omega^a\phi_a,f\}_{\rm PB}$, the corresponding quantum version is
\begin{equation}
	\delta_\omega F := [\omega^a\Phi_a,F]_M = i\hbar\{\omega^a\phi_a,F
	\}_{\rm PB} + \mbox{other terms}
\end{equation}
which a priori differs from the classical expression. The discrepancy between
the classical and the quantum ``gauge'' transformations show up in higher
order derivatives, which seems to suggest that the quantum transformations
are ``larger'', i.e., slightly less local than their classical counterparts.\\
Another point to check is whether the space of physical quantities is 
invariant under such transformations. Consider thus an element $A$ satisfying
$[\Phi_a,A]_M^+=0, ~\forall a$, when then wants to prove that a ``gauge''
transformation does not take us away from this subspace, i.e., $\delta_\omega
[\Phi_a,A]_M^+=0, ~\forall a$. We get
\begin{eqnarray}
	\delta_\omega ([\Phi_a,A]_M^+) &=& [\delta_\omega\Phi_a,A]_M^++
	[\Phi_a,\delta_\omega A]_M^+\nonumber\\
	&=& [\omega^b [\Phi_b,\Phi_c]_M,A]_M^+ + [\Phi_a,[\omega^b\Phi_b,A]_M
	]_M^+
\end{eqnarray}
By noting that the physicality condition implies $\Phi_a*A=-A*\Phi_a, 
~\forall a$, we can then rewrite this as
\begin{equation}
	\delta_\omega([\Phi_a,A]_M^+) = \omega^b\left([[\Phi_b,\Phi_a]_M,A]_M^+
	+[[\Phi_a,\Phi_b]_M,A]_M^+\right) = 0
\end{equation}
Hence the condition $0=[\Phi_a,A]_M^+$ {\em is} a reasonable quantum analogue 
of $\phi_a=0$, as we had anticipated, since it fulfills the only two 
requirements on can make {\em a priori}, namely that it has the correct 
classical limit (the correspondence principle) and that it is invariant under
(quantum) gauge transformations. Notice that this even holds in the case 
where the structure coefficients depend upon the phasespace variables, as e.g.
in gravity.\\
For completeness we recall how Wigner functions are formed. In 
standard quantum mechanics in $d$ dimensions one defines \cite{wigner}
\begin{equation}
	W_\psi(q,p) = \int \psi^\dagger(x+\frac{1}{2}y)\otimes\psi(x-
	\frac{1}{2}y) e^{-iyp}\frac{dy}{(2\pi)^d}  
\end{equation}
where $\psi$ is some wave function. This definition is appropriate in
flat phasespaces and in the absence of external gauge fields. The case of
gauge fields, however, can be treated by defining $\psi(x\pm\frac{1}{2}y)$ in
a covariant manner using parallell transport, \cite{EGV}. This can also
be extended to quantum theory in curved spacetimes, \cite{wletter}. More
generally, one defines $W_\psi$ by means of a {\em Wigner-Weyl-Moyal (WWM) 
map} as
\begin{equation}
	W_\psi = {\rm Tr}(\Pi(q,p) |\psi\rangle\langle\psi|)
\end{equation}
or for mixed states in terms of a density matrix $\rho$ as $W_\rho={\rm Tr}(\Pi
(q,p)\rho(q,p))$,
where $\Pi$ encodes all of the deformation information, \cite{wlie}, i.e.,
an operator $\hat{A}$ gets mapped to a phasespace function $A_W$ (its {\em
Weyl symbol})
\begin{equation}
	A_W = {\rm Tr}(\Pi \hat{A})
\end{equation}
and where the deformed product is given by
\begin{equation}
	A_W*B_W = {\rm Tr}(\Pi \hat{A}\hat{B}) = (\hat{A}\hat{B})_W
\end{equation}
and so on. It has been proven that such a \small{WWM}-map $\Pi$ exists for
a very large range of algebraic structures, \cite{wlie}.\\
One of the main features of the \small{WWM} formalism is that states and
observables are treated on an equal footing, in fact is given by a ``projection
operator'' in ${\cal A}_\hbar$, i.e., an $*$-idempotent $P$, $P*P=P$. Thus
${\cal C}_{\rm phys}\subseteq {\cal A}_\hbar$.\\
This can be clarified a bit by noting that the set of possible density
matrices $\rho$, and hence of possible Wigner functions, is the (closed) convex
hull of the set of $*$-idempotents (projections),
\begin{equation}
	{\cal C} = \overline{\rm co}\{P\in{\cal A}_\hbar~|~ P*P=P\}
\end{equation}
and consequently, ${\cal C}_{\rm phys}\subseteq{\cal C}$.\\
As a final comment worth making, we could note the possibility of letting the
quantum constraints be the classical ones but with the usual product of
phasespace variables replaced by the twisted one in a symmetric manner, i.e.,
replace $qp$ by $\frac{1}{2}(q*p+p*q)$. But when these are Darboux coordinates,
i.e., when they are canonically conjugate, then $\frac{1}{2}(q*p+p*q)$ and $qp=
pq$ are identical, since $q*p=qp+i\hbar, p*q=qp-i\hbar$. It is only when we
are working with a phasespace which is not covered by a global Darboux
coordinate patch that this possibility have any relevance. Since the 
non-flatness of the phasespace can always be absorbed into an appropriate set
of constraints on a larger, flat phasespace, this possibility is only of
academic interest. 

\section{Overcoming the Various Problems}
At the classical level, a number of problematic features can be present. The
constraints may not all be first class or the constraints might not be in
involution with the Hamiltonian, in both cases time evolution will take one
away from the original constraint surface. We will show how to overcome these
problems within the formalism of deformation quantisation. More precisely,
we want to show that provided one is willing to allow negative powers of
$\hbar$ (which is standard in the mathematical treatment of deformation 
quantisation), one can remove at the quantum level most if not all of the 
problems present at the classical level. It is this which is the main result
of this paper.\\
Problems may also arrise after a naive attempt at quantisation, such problems
will be referred to as {\em anomalies}, even though they may not be 
just the familiar
cases of central extensions of the algebra (a.k.a. Schwinger terms in current
algebra) but can encompass a wider range of problems. These too can be overcome
at least in some cases as will be shown.

\subsection{Second Class Constraints}
In Dirac's terminology, second class constraints are constraints $\psi_a$
which do not have weakly vanishing Poisson brackets, i.e.,
\begin{equation}
	\{\psi_a,\psi_b\}_{\rm PB} = C_{ab}\not\approx 0 \qquad \det C\neq 0
\end{equation}
where being weakly zero means vanishing when the constraints have been
taken into account (i.e., vanishing on the constraint surface). Dirac
originally
dealt with second class constraints by defining a new bracket, the {\em Dirac
bracket}, 
\begin{equation}
	\{f,g\}_{\rm DB} = \{f,g\}_{\rm PB} -\{f,\psi_a\}_{\rm PB} 
	(C^{-1})_{ab}\{g,\psi_b\}_{\rm PB} \label{eq:Dir}
\end{equation}
which then have the wonderful property
\begin{equation}
	\{\psi_a,\psi_b\}_{\rm DB} = 0
\end{equation}
i.e., the bracket of two second clas constraints become {\em strongly} zero.\\
There are a few problems with the Dirac bracket, however, most notably that
it is usually very hard to compute explicitly. As far as deformation 
quantisation is concerned though, one can simply replace Poisson brackets by
Dirac ones in the definition of the Moyal bracket to get a modified Moyal
bracket. We will not pursue that line of inquiery any further here, since this
approach will suffer even more from the calculational difficulties of the
Dirac bracket. Instead we will ask two other questions: (1) is it possible
to define ``first class'' quantum constraints $\Psi_a(\hbar)$ with $\Psi_a 
= ...+\psi_a+\hbar \Psi_a^{(1)}+...$ (we will have to include negative powers 
of $\hbar$
in order to be able to cancel the $C_{ab}$ from the zero'th order term), and
(2) can the Dirac bracket itself be seen as a deformation of the Poisson
bracket?\\
Let us first look for quantum constraints $\Psi_a$ satisfying
\begin{equation}
	[\Psi_a,\Psi_b]_M = i\hbar e_{ab}^c\Psi_c
\end{equation}
where we explicitly want to be able to have $e_{ab}^c\equiv 0$ too. Since we
now have to include negative powers of the deformation parameter $\hbar$, we
write
\begin{equation}
	\Psi_a = \sum_{n\in\mathbb{Z}}\hbar^n\Psi_a^{(n)}\qquad
	e_{ab}^c = \sum_{n=1}^\infty \hbar^n \stackrel{(n)}{e}_{ab}^c
\end{equation}
and compute the Moyal bracket. The $\hbar^0$ part of this becomes
\begin{equation}
	\sum_{n=1}^\infty \stackrel{(n)}{e}_{ab}^c\Psi_c^{(-n-1)}=
	\sum_{n\in\mathbb{Z}}\{\Psi_a^{(n)},\Psi_b^{(-n)}\}_{\rm PB}
	+\sum_{k=1}^\infty\sum_{n\in\mathbb{Z}}\omega_k(\Psi_a^{(n)},
	\Psi_b^{(-k-n)})
\end{equation}
which is then one of a family of equations the new constraints have to 
satisfy.\\
It will often be enough to assume $\Psi_c^{(n)} = 0, n=-2,-3,-4,...$, i.e.,
that only one negative power of the deformation parameter occurs. In this 
instance the set of equations read
\begin{eqnarray}
	0&=& \{\psi_a^{(-1)},\psi_b^{(-1)}\}_{\rm PB},\\
	0&=& \{\psi_a^{(-1)},\psi_b\}_{\rm PB} +
	\{\psi_a,\psi_b^{(-1)}\}_{\rm PB},\\
	\stackrel{(1)}{e}_{ab}^c\psi_c^{(-1)} -C_{ab} &=& 
	\omega_3(\psi_a^{(-1)},
	\psi_b^{(-1)})+\{\psi_a^{(-1)},\psi_b^{(1)}\}_{\rm PB}
	+\{\psi_a^{(1)},\psi_b^{(-1)}\}_{\rm PB},\nonumber\\
\end{eqnarray}
and so on.  One would often like to take $e_{ab}^c\equiv 0$ but this might not
always be possible, and not all choices of structure coefficients may be 
allowed. There will, however, be quite a lot of freedom in the particular
choice of algebra in the general case. \\
As was the case for first class constraints, we can introduce a cohomology
theory for this set of equations. Define
\begin{equation}
	\hat{B}_{ab}^{c_1} f_{c_1...c_n} = \sum_{\sigma \in S_n}
	{\rm sign}(\sigma) \{\psi_{\sigma(a)}^{(-1)},f_{\sigma(b)...
	\sigma(c_n)}\}_{\rm PB}
\end{equation}
then $\hat{B}^2=0$. Notice that this is the same cohomology construction
as for first class constraints, {\em except} that we now use the $\hbar^{-1}$
components and not the semiclassical ones for the definition of the
differential. With this the conditions read
\begin{eqnarray}
	\hat{B}_{ab}^c\psi_c^{(-1)} &=& 0\\
	\hat{B}_{ab}^c\psi_c &=& 0\\
	\hat{B}_{ab}^c\psi^{(1)} &=& \stackrel{(1)}{e}_{ab}^c\psi_c^{(-1)}
	+C_{ab}-\omega_3(\psi_a^{(-1)},\psi_b^{(-1)})
\end{eqnarray}
So, contrary to the first class case, both the the $\hbar^{-1}$ and the 
$\hbar^0$ terms are closed. Let
\begin{equation}
	\bar{Z}^p = {\rm Ker} \hat{B} \qquad \bar{B}^p = {\rm Im}\hat{B}
	\qquad \bar{H}^p=\bar{Z}^p/\bar{B}^p
\end{equation}
then
\begin{equation}
	\psi_a^{(-1)},\psi_a \in \bar{Z}^1
\end{equation}
but
\begin{equation}
	\psi^{(1)} \in \bar{Z}^1+\omega_3(\bar{Z}^1,\bar{Z}^1)
\end{equation}
which shows how the subsequent terms are build up from cohomological
ingredient at lower levels, i.e., that the $\hbar$-term is a closed form
plus the third order bracket, $\omega_3$, of two closed forms. Hence, as for
the case of first class constraints, the solutions are described by a 
cohomology. For second class constraints, however, the algebra of constraints 
is somewhat different. Instead of simply the algebra $\mathfrak{g}\subseteq 
{\cal A}=C^\infty(\Gamma)$ of classical
constraints, $\psi_a$, we have an extended algebra $\tilde{\mathfrak{g}}
\subseteq
{\cal A}=C^\infty(\Gamma)$, which is
an extension of $\mathfrak{g}$ by an Abelian algebra $\mathfrak{g}_0$ 
spanned by the 
$\hbar^{-1}$ components. The algebra $\mathfrak{g}_0$ defines a cohomology, 
$\bar{H}^p$, whose differential operator, $\hat{B}$, extends to all of the
extended algebra $\tilde{\mathfrak{g}}$, making the situation somewhat more
difficult. The extension $\tilde{\mathfrak{g}}$ need not be a central 
extension,
since the condition $\hat{B}^c_{ab}\psi_c=0$ does not imply $\{\psi_a,
\psi_b^{(-1)}\}_{\rm PB}=0$. It is clear, however, that central extensions 
would be a good starting point when one wants to construct $\tilde{
\mathfrak{g}}$.
This once more brings us back to the Lie algebra cohomology of 
$\mathfrak{g}$, since
the second cohomology class thereof describes the various possible central
extensions. We will not elaborate more on the cohomological element here.\\
It is important to emphasise that the desire to remove second class constraints
{\em forces} the pressence of negative powers of $\hbar$ upon us. But provided
one is willing to pay this small price, one can treat second and first class
constraints in the same manner {\em at the quantum level.} This of course 
implies that one has the redefine the classical limit slightly, a point to
which we will return towards the end of this paper.\\
In the following subsection we will consider a particular example of a
physical system with second class constraints and show how to use
deformation quantisation explicitly. But first we will return to the Dirac
bracket construction again.\\
The Dirac bracket, (\ref{eq:Dir}), can be seen as a deformation of the
classical Poisson bracket in its own right. The deformation parameter is
in this case a matrix valued function, namely $C_{ab}$ itself. This is most
easily seen in the following simple toy model where we have an algebra of
first class constraints $\phi_a$ and a pair of second class ones, $\psi,
\bar{\psi}$, satisfying the generic algebra
\begin{eqnarray*}
	\{\phi_a,\phi_b\}_{\rm PB} &=& c_{ab}^c\phi_c\\
	\{\phi_a,\psi\}_{\rm PB} &=& c_a^b\phi_b\\
	\{\phi_a,\bar{\psi}\}_{\rm PB} &=& \bar{c}_a^b\phi_b\\
	\{\psi,\bar{\psi}\}_{\rm PB} &=& k
\end{eqnarray*}
The Dirac bracket of the first class constraints then reads
\begin{displaymath}
	\{\phi_a,\phi_b\}_{\rm DB} = c_{ab}^c\phi_c-k^{-1}c_a^c\bar{c}_b^d
	\phi_c\phi_d
\end{displaymath}
One can either see this as defining a ``quadratic algebra'' or as a deformation
of the Poisson bracket. This latter point is particularly clear when one
considers two arbitrary functions $f,g$ and compute their Dirac bracket
\begin{displaymath}
	\{f,g\}_{\rm DB} = \{f,g\}_{\rm PB} -k^{-1}\{\psi,f\}_{\rm PB}
	\{\bar{\psi},g\}_{\rm PB} := \{f,g\}_{\rm PB} - k^{-1}\omega(f,g)
\end{displaymath}
Contrary to the Moyal bracket deformation of the classical Poisson brackets,
the Dirac bracket, however, does not involve higher order derivatives of the
observables. It is a deformation all the same; the limit $k\rightarrow\infty$
corresponding to the classical Poisson bracket. Contrary to the Moyal bracket,
the higher oder brackets do not involve higher order derivatives. Deformations
such that the $n$'th order bracket contains only $n$'th order derivatives
is known as a deformation of {\em Vey type}. Hence, the Moyal bracket is of
Vey type whereas the Dirac bracket is not.\\
The Dirac bracket amounts to replacing the Poisson bracket by a new first
order bracket
\begin{equation}
	\{f,g\}_2 = \alpha_{qq}(C)\frac{\partial f}{\partial q}
	\frac{\partial g}{\partial q} -\beta_{pq}(C)\frac{\partial f}
	{\partial p}\frac{\partial g}{\partial q} -\beta_{qp}(C)
	\frac{\partial f}{\partial q}\frac{\partial g}{\partial p}
	+\alpha_{pp}(C)\frac{\partial f}{\partial p}\frac{\partial g}
	{\partial p}
\end{equation}
where
\begin{eqnarray}
	\alpha_{pp}(C) = O(C) &\qquad & \alpha_{qq} = O(C)\\
	\beta_{pq}(C) = -1+O(C) &\qquad& \beta_{qp} = +1+O(C)
\end{eqnarray}
where $C$ is allowed to be a matrix valued function of the phasespace
variables. We will use the name {\em secondary deformations} for such
deformations. A priori, such secondary brackets will not define Lie algebras;
one can only ensure this to $O(C)$. \footnote{This is somewhat similar to the 
situation in equivariant cohomology -- if one attempts to use the secondary 
bracket $\{\cdot,\cdot\}_2$ to define a cohomology theory, i.e., a derivation
$s$, one will get $s^2=O(C)$ in the generic case, and one would not get a
standard cohomology theory.}\\
Before we move on to consider some examples and to study anomalies, it is worth
recalling that other ways of removing second class constraints exist. These
are inspired by the \small{BRST} approach to quantisation, \cite{brst}. One
enlarges the phasespace by including Grassmann odd variables, $\eta, {\cal 
P}$, and then replaces the constraints $\psi_i$ by $\Psi_i = \psi_i+\sum_{n=1}
\psi_i^{(n)}\underbrace{\chi...\chi}_n$, where $\chi$ can be either 
$\eta$ or $\cal P$.
The extra Grassmann variables (the ghosts) here correspond to our 
deformation parameter $\hbar$, and is just one indication of a connection
between \small{BRST} and deformation quantisation. We will comment a bit more
on this connection in the section on the underlying geometrical structure.

\subsubsection{Example: $O(N)$ Non-Linear $\sigma$-Model}
Consider an $O(N)$-valued field $\phi$, and take the Lagrangean to be
\begin{equation}
	{\cal L} = \frac{1}{2}\partial_\mu\phi^A\partial^\mu\phi_A
	+\frac{1}{2}\lambda(\phi^A\phi_A-1)
\end{equation}
where $\mu$ is a (flat) spacetime index and $A$ is the $O(N)$-index. In the
Hamiltonian formalism of this $O(N)$ non-linear $\sigma$-model, we have the
following constraints
\begin{eqnarray} 
	\psi_1 &=& \phi_A\phi^A-1\\
	\psi_2 &=& \pi^A\phi_A
\end{eqnarray}
where $\pi_A$ is the canonically conjugate momentum of $\phi_A$. The 
constraints are second class and satisfy
\begin{equation}
	\{\psi_1(x),\psi_2(x')\}_{\rm PB} = 2\delta(x-x')\phi_A\phi^A
\end{equation}
which is clearly non-zero on the constraint surface.\\
We now want to find quantum modified constraints $\Psi_i,i=1,2$ with vanishing
Moyal brackets and with 
\begin{equation}
	\Psi_i = \hbar^{-1}\psi_i^{(-1)}+\psi_i
\end{equation}
This leads to the following set of equations (suppressing the spacetime
variables $x,x'$ and the resulting Dirac $\delta$-functions)
\begin{eqnarray}
	\{\psi_i^{(-1)},\psi_j^{(-1)}\}_{\rm PB} &=& 0\\
	\{\psi_i^{(-1)},\psi_j\}_{\rm PB} + \{\psi_i,\psi_j^{(-1)}\}_{\rm PB}
	&=& 0\\
	\omega_3(\psi_i^{(-1)},\psi_j^{(-1)}) &=& \{\psi_i,\psi_j\}_{\rm PB}
	= 2\epsilon_{ij}\phi_A\phi^A\\
	\omega_n(\psi_i^{(-1)},\psi_j^{(-1)}) &=& 0 \qquad n\geq 5
\end{eqnarray}
We will make the following {\em Ansatz} for the quantum part of the 
constraints
\begin{equation}
	\psi_1^{(-1)} = \psi_1 f(\phi) \qquad \psi_2^{(-1)} = \psi_2 g(\phi,
	\pi)
\end{equation}
This yield the following set of coupled
equations
\begin{eqnarray}
	0&=& 2 \phi^2 fg+\psi_1 g \phi\cdot\frac{\partial f}{\partial\phi}
	+\psi_1\psi_2\frac{\partial f}{\partial\phi}\cdot\frac{\partial g}
	{\partial\pi} + 2 f\psi_2\phi\cdot\frac{\partial g}{\partial\pi}\\
	0&=&2\phi^2 (f+g)+2\psi_2\phi\cdot\frac{\partial g}{\partial\pi}
	+\psi_1\phi\cdot\frac{\partial f}{\partial\phi}\\
	0&=&\left(2\delta_{AB}\frac{\partial f}{\partial\phi_C}+2\phi_A
	\frac{\partial^2 f}{\partial\phi_B\partial\phi_C}+\frac{1}{3}
	(\phi^2-1)\frac{\partial^3 f}{\partial\phi_A\partial\phi_B
	\partial\phi_C}+\mbox{cyclic}\right)\times\nonumber\\
	&&\left(\phi^A\frac{\partial^2 g}{\partial\pi^B\partial\pi^C}
	+\frac{1}{3}\phi\cdot\pi \frac{\partial^3 g}{\partial\pi^A
	\partial\pi^B\partial\pi^C}+\mbox{cyclic}\right)-2\phi^2
\end{eqnarray}
where {\em cyclic} denotes a sum over cyclic permutations of $A,B,C$.\\ We can
ensure $\omega_n(\Psi_1,\Psi_2)\equiv 0, n\geq 5$ if we take $g$ to be only
quadratic in the momenta, i.e.,
\begin{equation}
	g(\phi,\pi) = \alpha_{AB}\pi^A\pi^B+\beta_A\pi^B+\gamma
\end{equation}
where $\alpha_{AB},\beta_A,\gamma$ can depend on $\phi$. Notice, furthermore,
that the $\phi$-derivatives of $g$ do not appear at all in this set of
quations, hence there is a lot of freedom in choosing the $\phi$-dependency
of $g$. We will use this to assume $\alpha,\beta$ to be independent of 
$\phi$.\\
By combining the first two equations we get 
\begin{eqnarray}
	2\phi^2 fg + 2f(\phi\cdot\pi)\phi^A(\alpha_{AB}\pi^B+\beta_A) &=&
	-f(\phi^2-1)\phi\cdot\partial f\\
	(g-f)\phi\cdot\partial f + (\phi\cdot\pi)(\alpha_{AB}\pi^B+\beta_A)
	\partial^Af &=& 0
\end{eqnarray}
where we have written $\partial f$ for $\frac{\partial f}{\partial\phi}$.
Collecting powers of the momentum we get
\begin{eqnarray}
	\alpha_{AB}\pi^A\pi^B\phi^C\partial_Cf &=& -\alpha_{BC}\pi^A\pi^B\phi^C
	\partial_Af\\
	\beta_A\phi^B\pi_A\partial_Bf &=& -\beta_A\phi^B\pi_B\partial_Af\\
	(\gamma-f)\phi^A\partial_Af &=& 0
\end{eqnarray}
To get an understanding of these equations we will consider first the case of
$N=1$. The third condition then implies $\gamma=f$. Using this we arrive at
\begin{equation}
	2 f'+2\phi f'' + (\phi^2-1) f''' = 2\phi/\alpha \label{eq:f}
\end{equation}
The form of this equation suggests an expansion on Hermite polynomials of 
$h=f'$. Thus, write
\begin{equation}
	h = \sum_{n=0}^\infty c_nH_n(\phi)
\end{equation}
Inserting this in (\ref{eq:f}) we get
\begin{equation}
	\sum_{n=0}^\infty\left[c_n+2\phi(n+1)c_{n+1} + 2(n+1)(n+2)(\phi^2-1)
	c_{n+2}\right]H_n = \frac{1}{2\alpha} H_1
\end{equation}
where we have used the explicit form for $H_1$ (i.e., $H_1(x)=2x$) and the
standard formula
\begin{displaymath}
	\frac{d}{dx}H_n(x) = 2n H_{n-1}(x)
\end{displaymath}
Now, multiplying by $H_1 e^{-\phi^2}$ and integrating over $\phi$, we get
from the orthonormality of the Hermite polynomials
\begin{equation}
	2\sqrt{\pi}\left[ c_1+12 c_3 - 12 c_3+480 c_5
	+72 c_3\right] = \frac{\sqrt{\pi}}{\alpha} \label{eq:cc}
\end{equation}
From this we get
\begin{eqnarray}
	h&=&c_1 H_1+c_3 H_3+c_5 H_5\\
	&=& (2c_1-12c_3+120c_5) \phi+(8c_3-160c_5)\phi^3+32c_5\phi^5
\end{eqnarray}
where the coefficients $c_1,c_3,c_5$ are restricted by (\ref{eq:cc}). There 
are two undetermiend coefficients since the equation defining $h$ is of
second order. The function $f$ is then the primitive of $h$, i.e.,
\begin{equation}
	f(\phi) = c_0+(c_1-6c_3+60c_5) \phi^2+
	(2c_3-40c_5)\phi^4+\frac{16}{3}c_5\phi^6
\end{equation}
where $c_0$ is an undetermined constant. Note, by the way, that, on the 
constraint surface, $f=const$. Hence our new quantum constraints are
\begin{eqnarray}
	\Psi_1 &=& \psi_1\left(1+\hbar^{-1}f(\phi)\right)\\
	\Psi_2 &=& \psi_2\left(1+\hbar^{-1}(\alpha \pi^2+\beta\pi+
	f(\phi))\right)
\end{eqnarray}
We have now constructed a set of quantum constraints $\Psi_1,\Psi_2$ 
satisfying an abelian algebra under the Moyal bracket,
\begin{equation}
	[\Psi_1,\Psi_2]_M=0
\end{equation}
and the second class constraints have consequently been ``lifted'' to an
abelian, but deformed, algebra. This illustrates the general procedure of
deformation quantisation of Hamiltonian systems with second class constraints.
For the remaining part of this paper, we will implicitly assume that second
class constraints have been dealt with in this manner, and will thus
concentrate on first class constraints.\\
The case $N>1$ is treated in an analogous manner, but the solution will here
be in terms of many-variable Hermite polynomials, $H_{n_1,...,n_N}(\phi_1,...
,\phi_N)$ which satisfy similar recursion relations. Unfortunately, the
explicit expressions quickly  become highly complicated and messy and we will
not give them here. It is, however, fairly straighforward to do so -- it
merely corresponds to finding solutions of the many-dimensional harmonic
oscillator Schr\"{o}dinegr equation. In a multi index notation, moreover,
the equations essentialy reduce to the $N=1$ case (with a few subtleties).

\subsection{Constraints not in Involution with the Hamiltonian}
Another problem which may occur is the failure of the constraints to be in
involution with the Hamiltonian, i.e.,
\begin{equation}
	\{h,\phi_a\}_{\rm PB} \neq V_a^b \phi_b
\end{equation}
In this instance one cannot simply impose the constraints for some initial
values of the phasespace variables, but have to impose them at each time
step, since time evolution does not preserve the constraints.\\
Explicitly, we will assume the constraints to be first class but satisfying
\begin{equation}
	\{h,\phi_a\}_{\rm PB} = V_a^b\phi_b + \chi_a
\end{equation}
where $\chi_a\not\approx 0$. The Poisson bracket of $\chi_a$ with the 
constraints will be denoted by $\lambda_{ab}$,
\begin{equation}
	\{\phi_a,\chi_b\}_{\rm PB} = \lambda_{ab}
\end{equation}
with $\lambda_{ab}\not\approx e_{ab}^c\chi_c$ for some function $e_{ab}^c$.
This implies that $\chi_a$ cannot be seen as merely a new first class
constraint.\\
The task is to find new quantum constraints and Hamiltonian such that
\begin{equation}
	[H,\Phi_a]_M = i\hbar \tilde{V}_a^b\Phi_b
\end{equation}
with $\tilde{V}_a^b=V_a^b+O(\hbar)$. We will again assume $V_a^b$ to be 
independent of the phasespace coordinates.\\
Inserting an expansion in $\hbar$,
\begin{eqnarray*}
	H &=& \sum_n \hbar^n H_n\\
	\Phi_a &=& \sum_n \hbar^n \Phi_a^{(n)}\\
	\tilde{V}_a^b &=& \sum_n \hbar^n \stackrel{(n)}{V_a^b}
\end{eqnarray*}
we get for the $\hbar^0$ part
\begin{equation}
	\sum_{k,l,n} \delta_{k+l,-2n}\omega_{2n+1}(H_k,\Phi_a^{(l)})
	=\sum_n \stackrel{(n)}{V_a^b}\Phi_b^{(-n)}
\end{equation}
which clearly shows that negative powers of $\hbar$ are needed -- otherwise
we would get the condition $V_a^b\phi_b+\chi_a = \stackrel{(0)}{V_a^b}\phi_b$
which is impossible. It does, however, make sense to assume that $H$ and 
$\tilde{V}_a^b$ have only non-negative powers of the deformation parameter,
whereas the only negative powers of $\hbar$ appear in $\Phi_a$. We can even
assume that only powers $\hbar^n, n\geq -1$ occur.\\ 
With this assumption, the condition coming from the $\hbar^{-1}$ part reads
\begin{equation}
	\omega_1(h,\Phi_a^{(-1)}) = \stackrel{(0)}{V_a^b}\Phi_b^{(-1)}
	\label{eq:inv}
\end{equation}
i.e., the corresponding term in the quantum constraints {\em are} in involution
with the Hamiltonian, with the structure coefficients being given by the
involutive part of the classical relation, i.e., $\stackrel{(0)}{V_a^b} = 
V_a^b$.\\
The ``classical'' part, i.e., the $\hbar^0$ contribution, reads
\begin{equation}
	V_a^b\phi_b+\chi_a+\omega_1(H_1,\Phi_a^{(-1)})=
	\stackrel{(0)}{V_a^b}\phi_b+\stackrel{(1)}{V_a^b}\Phi_b^{(-1)}
\end{equation}
which simplifies upon putting $\stackrel{(0)}{V_a^b}=V_a^b$ as mentioned 
above.\\
Since there are too many indeterminates in this problem (the $H$, $\Phi_a$
and $\tilde{V}_a^b$), we will make a further simplifying {\em Ansatz}, namely
$H=h$, i.e., that the Hamiltonian receives no quantum corrections at this
level -- all quantum modifications will be put in the constraints and the
structure coefficients. With this condition we arrive at
\begin{equation}
	\stackrel{(1)}{V_a^b}\Phi_b^{(-1)} = \chi_a
\end{equation}
Together with the involutive requirement, (\ref{eq:inv}), this leads to
a differential equation for $\stackrel{(1)}{V_a^b}$, namely
\begin{equation}
	\{h,(\stackrel{(1)}{V^{-1}})_a^b\chi_b\}_{\rm PB} = V_a^b
	(\stackrel{(1)}{V^{-1}})_b^c\chi_c
\end{equation}
Letting $v_a^b=\stackrel{(1)}{V_a^b}$ and using a matrix notation, this
can be written as
\begin{equation}
	\{h,v^{-1}\}_{\rm PB}\chi + v^{-1}\{h,\chi_b\}_{\rm PB} = Vv^{-1}\chi
\end{equation}
which we can also write as
\begin{equation}
	vVv^{-1}\chi-\dot{\chi} = v\{h,v^{-1}\}_{\rm PB}\chi = -\{h,v\}_{\rm 
	PB} v^{-1}\chi \equiv -\dot{v}v^{-1}\chi
\end{equation}
In the simple case where we have only one constraint this equation becomes
\begin{displaymath}
	V-\frac{d}{dt}\ln\chi = -\frac{d}{dt}\ln v
\end{displaymath}
which has the solution
\begin{displaymath}
	v(t) =c_0e^{-Vt}\chi
\end{displaymath}
where we have used that $V$ is assumed independent of the phasespace 
coordinates in order to carry out the time integration.\\
Hence constraints which at the classical level aren't in involution with the
Hamiltonian can, by deformation quantisation, be turned into quantum 
constraints which are. The price being that the constraints have a singular
classical limit $\hbar\rightarrow 0$. This price turns out to be the standard
fare when dealing with classical constraints with problems, or when one
wants to eliminate problems arrising from a naive quantisation.
 
\subsection{Anomalies and their Lifting}
Upon deformation, the constraint algebra can be modified in a number of 
different ways. We will refer to such modification in general as {\em 
anomalies}. We have a natural hierarchy of such anomalies:
\begin{itemize}
	\item Anomaly of zero'th order: $[\phi_a,\phi_b]_M \neq i\hbar
	\{\phi_a,\phi_b\}_{\rm PB} = i\hbar c_{ab}^c\phi_c$.
	\item Anomaly of first order: $[\Phi_a,\Phi_b]_M \neq i\hbar
	c_{ab}^c\Phi_c$
\end{itemize}
The anomalies of zero'th order will in general be removable by deforming the
constraints. One particularly important situation is when
\begin{equation}
	\omega_n(\phi_a,\phi_b) = k_{ab}^{(n)} = const.
\end{equation}
the Moyal bracket of two undeformed constraints is then
\begin{equation}
	[\phi_a,\phi_b]_M = i\hbar c_{ab}^c\phi_c + k_{ab}
\end{equation}
with $k_{ab} = \sum_nk_{ab}^{(n)}\hbar^{2n+1}$. Hence we have a central 
extension of the original constraint algebra. This is the way an anomaly very
often shows up. We want to find deformed constraints, i.e., quantum 
constraints,
$\Phi_a$ such that the anomaly is ``lifted'', $[\Phi_a,\Phi_b]_M =i\hbar
\tilde{c}_{ab}^c\Phi_c$ -- where we allow for a change in algebra for 
$\hbar\neq 0$. One should note that this procedure is analogous to the way 
we ``lifted'' second class constraints and the way in which we handled 
constraints not in involution with the Hamiltonian; the procedure 
thus shows the great flexibility and power of deformation quantisation.\\ 
It will in general be sufficient to assume $\stackrel{(n)}{c_{ab}^c}=0$ for
$n<0$ and $\Phi_a^{(-k)}=0$ for $k\geq 2$. Furthermore, suppose $k_{ab}^{(n)}$
vanishes after a certain stage $N$, it is then natural to choose 
$\stackrel{(n)}{c_{ab}^c}=\Phi_a^{(n)}=0$ for $n\geq N$ too. Moreover, this 
will be the typical situation as the constraints usually are polynomial and
hence have vanishing $n$-order brackets for $n$ sufficiently large.\\
The general conditions become
\begin{eqnarray}
	\stackrel{(0)}{c_{ab}^c}\Phi_c^{(-1)} &=& \omega_1(\phi_a,
	\Phi_b^{(-1)})+\omega_1(\Phi_a^{(-1)},\phi_b)\\
	\stackrel{(0)}{c_{ab}^c}\phi_c+\stackrel{(1)}{c_{ab}^c}\Phi_c^{(-1)}
	&=& \omega_1(\Phi_a^{(-1)},\Phi_b^{(1)})+\omega_1(\Phi_a^{(1)},
	\Phi_b^{(-1)})+c_{ab}^c\phi_c\\
	\stackrel{(0)}{c_{ab}^c}\Phi_c^{(1)}+\stackrel{(1)}{c_{ab}^c}\phi_c
	+\stackrel{(2)}{c_{ab}^c}\Phi_c^{(-1)} &=& \omega_1(\phi_a,
	\Phi_b^{(1)})+\omega_1(\Phi_a^{(1)},\phi_b)+\omega_1(\Phi^{(2)},
	\Phi_b^{(-1)})+\nonumber\\
	&& \omega_1(\Phi_a^{(-1)},\Phi_b^{(2)})+
	\omega_3(\phi_a,\Phi_b^{(-1)})+\omega_3(\Phi_a^{(-1)},\phi_b)\nonumber
	\\
	&&\\
	\stackrel{(0)}{c_{ab}^c}\Phi_c^{(2)}+\stackrel{(1)}{c_{ab}^c}
	\Phi_c^{(1)}+\stackrel{(2)}{c_{ab}^c}\phi_c+\stackrel{(3)}
	{c_{ab}^c}\Phi_c^{(-1)} &=& \omega_1(\Phi_a^{(2)},\phi_b)+
	\omega_1(\phi_a,\Phi_b^{(2)})+\omega_1(\Phi_a^{(1)},\Phi_b^{(1)})
	+\nonumber\\
	&&\omega_1(\Phi_a^{(-1)},\Phi_b^{(3)})+\omega_1(\Phi_a^{(3)},\Phi_b
	^{(-1)})+k_{ab}^{(3)}
\end{eqnarray}
and so on.\\
We quickly recognize a cohomological element to these equations. In terms of
the aforementioned cohomology operators $\hat{A}_{ab}^c, \hat{B}_{ab}^c$, 
we can write the set of equations on the form
\begin{eqnarray}
	\stackrel{(0)}{c}_{ab}^c \Phi_c^{(-1)} &=& \hat{A}_{ab}^c\Phi_c^{(-1)}
	= \hat{B}_{ab}^c\phi_c\\
	(\stackrel{(0)}{c}_{ab}^c-c_{ab}^c)\phi_c + \stackrel{(1)}{c}_{ab}^c
	\Phi_c^{(-1)} &=& \hat{B}_{ab}^c\Phi_c^{(1)}\\
	\stackrel{(0)}{c}_{ab}^c\Phi_c^{(1)}+\stackrel{(1)}{c}_{ab}^c\phi_c
	+\stackrel{(2)}{c}_{ab}^c\Phi_c^{(-1)} &=& \hat{A}_{ab}^c\Phi_c^{(1)}
	+\hat{B}_{ab}^c\Phi_c^{(2)}+\omega_3(\phi_a,\Phi_b^{(-1)})+
	\omega_3(\Phi_a^{(-1)},\phi_b)\nonumber\\
\end{eqnarray}
etc., stating that the anomaly prevents $\Phi_a^{(-1)}$ from being 
$\hat{A}$-closed, or equivalently, $\phi_a$ from being $\hat{B}$-closed.
In this way, an anomaly is recognized as an obstruction, and hence as
corresponding to a non-trivial cohomology class.\\
Let us again for simplicity restrict ourselves to the simplest possible
non-Abelian algebra $c_{12}^1=1$, $a,b,c=1,2$. We will then write simply
$k$ for $k_{12}$. We can furthermore assume $\omega_{2n+1}(\phi_a,\phi_b)=0$
for $n\geq 2$, i.e., $N=2$ (a general value for $N$ can be treated similarly
but the notation quickly becomes cluttered). 
This implies we can take $\stackrel{(n)}{c_{ab}^c}=
\Phi_a^{(n)}=0, n\geq 2$ too. Moreover, it is natural in any case to pick
$\stackrel{(0)}{c_{ab}^c}=c_{ab}^c$. The full set of conditions read
\begin{eqnarray}
	\Phi_2^{(-1)} &=& \omega_1(\Phi_1^{(-1)},\phi_2)+\omega_1(\phi_1,
	\Phi_2^{(-2)})\\
	\stackrel{(1)}{c_{12}^c}\Phi_c^{(-1)} &=& \omega_1(\Phi_1^{(-1)},
	\Phi_2^{(1)})+\omega_1(\Phi_1^{(1)},\Phi_2^{(-1)})\\
	\Phi_2^{(1)}+\stackrel{(1)}{c_{12}^c}\phi_c &=& \omega_1(\phi_1,
	\Phi_2^{(1)})+\omega_1(\Phi_1^{(1)},\phi_2)+\omega_3(\phi_1,
	\Phi_2^{(-1)})+\nonumber\\
	&&\qquad \omega_3(\Phi_1^{(-1)},\phi_2)\\
	\stackrel{(1)}{c_{12}^c}\Phi_c^{(1)} &=& \omega_1(\Phi_a^{(1)},
	\Phi_2^{(1)})+k\\
	\omega_3(\phi_1,\Phi_2^{(1)})+\omega_3(\Phi_a^{(1)},\phi_2)&=&0
\end{eqnarray}
It turns out, that a solution can be found if we make the {\em Ans\"{a}tze}
\begin{equation}
	\Phi_a^{(1)} =\alpha_a^b\phi_b+\beta_a\qquad 	
	\Phi_a^{(-1)} = \gamma_a^b\phi_b
\end{equation}
where $\alpha,\beta,\gamma$ are constants. The appearance of a $\beta$ is 
needed because of the fourth relation above would otherwise yield $k\propto
\phi$ which is explicitly assumed not to be the case. Furthermore, the first
relation shows that $\Phi_a^{(-1)}$ can only be proportional to $\phi_a$ and
not be linearly dependent upon it.\\
By inserting the {\em Ans\"{a}tze} in the relations above, one gets
\begin{eqnarray*}
	\gamma_2^1 &=& \gamma_1^1= 0\\
	\stackrel{(1)}{c_{12}^1}\gamma_1^2 &=& \alpha_2^1\gamma_1^2
	+\alpha_1^1\gamma_2^2\\
	{\rm Tr}\alpha &=& \alpha_2^2+\stackrel{(1)}{c_{12}^2}\\
	\alpha_2^1+\stackrel{(1)}{c_{12}^1} &=& 0\\
	\beta_2 &=& \gamma_2^2k\\
	\det\alpha &=& \stackrel{(1)}{c_{12}^c}\alpha_c^2\\
	\stackrel{(1)}{c_{12}^c}\alpha_c^1 &=& 0\\
	\stackrel{(1)}{c_{12}^c}\beta_c &=& k
\end{eqnarray*}
From which we get the structure coefficients
\begin{equation}
	\stackrel{(1)}{c_{12}^c}=\left\{\begin{array}{ll}
	-\alpha_2^1 &c=1\\
	\alpha_1^1 &c=2\end{array}\right.
\end{equation}
and the linear part, $\beta$, to be
\begin{equation}
	\beta_1 = \left(\gamma_2^2-\frac{1}{\alpha_1^1}\right) k \qquad
	\beta_2=\gamma_2^2 k
\end{equation}
Which shows, that it is the central extension, $k$, which necessitates the
linear part $\beta$. Finally we get for the matrices $\alpha,\gamma$ that
they must be of the form
\begin{equation}
	\alpha = \left(\begin{array}{cc} \alpha_1 & -\alpha_1\\
	\alpha_2 & -\alpha_1\end{array}\right)\qquad
	\gamma=\left(\begin{array}{cc} 0 & \gamma_1\\ 0 & \gamma_2
	\end{array}\right)
\end{equation}
subject to
\begin{equation}
	-2\alpha_2\gamma_1 = \alpha_1\gamma_2
\end{equation}
but otherwise they can be chosen at random (but non-zero, otherwise $\gamma
\equiv 0$ or $\alpha\equiv 0$ which is clearly not what we want).\\
Explicitly, the new quantum constraints read
\begin{eqnarray}
	\Phi_1 &=& \hbar^{-1}\gamma_1\phi_2+\phi_1+\hbar\alpha_1(\phi_1-
	\phi_2)+\hbar(\gamma_2-\alpha_1^{-1})k\\
	\Phi_2 &=& \hbar^{-1}\gamma_2\phi_2+\phi_2+\hbar(\alpha_2\phi_1-
	\alpha_1\phi_2+\gamma_2k)
\end{eqnarray}
and the Moyal algebra is
\begin{equation}
	[\Phi_1,\Phi_2]_M = (1-\alpha_2\hbar)\Phi_1+\alpha_1\hbar\Phi_2
\end{equation}
which is still isomorphich to the classical, undeformed Poisson algebra, since
that was the unique non-Abelian algebra with two generators. In general the
Moyal and the Poisson algebra needn't be isomorphic, however, the latter only 
being a limit (or contraction), $\hbar\rightarrow 0$, of the former.\\
The important thing to learn from this example is that the quantum constraints
acquire quantum modifications of negative order in $\hbar$ as well as of 
positive orders, and that, furthermore, the structure coefficients too get
quantum modified, but only with positive powers of $\hbar$. Note, by the way,
that it would be inconsistent to choose $\tilde{c}_{ab}^c=c_{ab}^c$, hence
anomalies of the zero'th order can be lifted, provided one is willing to 
modify the structure coefficients.\\
In the case where $k_{ab}$ is not a constant, we will have to let $\alpha,
\beta,\gamma$ too depend on the phasespace variables. In this way, the 
conditions on these coefficient become differential equations, just like they
did for the case of constraints not in involution with the Hamiltonian, and we
will not comment further on that here.\\
Anomalies of first order can result from a bad choice of quantum constraints,
but if that is not the case, they cannot in general be lifted. It is difficult
to think of any examples of such systems; they would correspond to a set-up in
which the only consistent quantisations were drastically different from their
classical counterparts. If the Moyal algebra becomes simply a non-linear
algebra, i.e., if for instance
\begin{displaymath}
	[\Phi_a,\Phi_b]_M = i\hbar c_{ab}^{~~c}\Phi_c+i\hbar^2d_{ab}^{~~cd}
	\Phi_c\Phi_d
\end{displaymath}
then we can lift this anomaly by extending the constraint algebra to include
$\Phi_{ab}:=\hbar \Phi_a\Phi_b$. Typically, this would lead to an infinite
dimensional constraint algebra.\\
Hamachi, \cite{Hamachi}, has recently considered another kind of anomaly
``smoothening''. Again, it turns out that anomalous contributions can be
removed at the Moyal algebra level. However, Hamachi restricts himself to
systems with constraints only depending on momentum. The set-up presented
here is much more general, although our results are not as rigorous as his.
Just like in our case, the anomaly reappears as a troublesome $\hbar
\rightarrow 0$ limit.

\section{Examples: Yang-Mills and Gravity}
Possibly the two most important kind of theories in modern theoretical physics
are Yang-Mills theories and General Relativity together describing all known
forces in the universe. It is consequently of interest to devote some time
to the study of their deformation quantisation. Especially when one considers
the important differences in the structure of their respective constraint
algebras. Where the structure coefficients of Yang-Mills theory are independent
of the fields (thereby forming a true Lie algebra), the corresponding
coefficients of the gravitational case do depend on the fields (the metric
to be precise). Furthermore, at least two set of constraint algebras exist
for general relativity, the \small{ADM} ones and the Ashtekar variables, the
latter formally resembling Yang-Mills theory quite a bit. Recently, a third
algebra of constraints for gravity has been proposed, where the algebra is
of the form $C^\infty(\Sigma)\ltimes{\rm Diff}(\Sigma)$ where $\Sigma$ is
the Cauchy hypersurface and $\ltimes$ denotes semidirect product. We will 
return to these after having dealt with Yang-Mills theory.

\subsection{Yang-Mills Theory}
For a Yang-Mills theory with some gauge algebra $\mathfrak{g}$, the 
constraints are
\begin{equation}
	{\cal G}_a = D_i\pi^i_a
\end{equation}
where $\pi^i_a$ is the momentum conjugate to $A_i^a$, $D_i$ denotes the
gauge covariant derivative and $i$ is a space index (which we will take to run
from one to three) and $a$ is a Lie algebra index. The constraint algebra is
\begin{equation}
	\{{\cal G}_a(x),{\cal G}_b(x')\}_{\rm PB} = c_{ab}^c\delta(x-x'){\cal 
	G}_c(x)
\end{equation}
where $c_{ab}^c$ are the structure coefficients of $\mathfrak{g}$.\\
Since the constraints are quadratic in the variables (they have the form
${\cal G} \sim \partial\pi + \pi A$), we can take ${\cal G}_a$ to be the
quantum constraints too, i.e.,
\begin{equation}
	[{\cal G}_a(x),{\cal G}_b(x')]_M = i\hbar c_{ab}^c\delta(x-x')
	{\cal G}_c(x)
\end{equation}
The last bracket is the one with the Hamiltonian
\begin{equation}
	{\cal H} = \frac{1}{2}\pi_a^i\pi_i^a - \frac{1}{2}F_{ij}^a F^{ij}_a
\end{equation}
Although $\cal H$ is fourth order in the fields (it has an $A^4$ term)
the constraints are merely quadratic and hence $\omega_n({\cal H},{\cal G}_a)
=0$ for $n\geq 1$. Thus we can take $\cal H$ to be also the quantum 
Hamiltonian, and we get no anomalies (i.e., no anomalies from the deformation
quantisation procedure, that is, other types of anomalies, especially global 
ones, are still possible).\\
The constraint equations become finite order functional differential equations
\begin{equation}
	0 = [{\cal G}_a,W]_M^+ = 2(D_i\pi^i_a) W+\frac{1}{4}i\hbar^2\delta^j_k
	c_{ab}^c\frac{\delta^2W}{\delta A^c_j\delta \pi_b^k}
	\label{eq:YMphys}
\end{equation}
where we have used $[f,g]_M^+=2f\cos(\frac{1}{2}\Delta) g$,\cite{Dahl}. We
have also suppressed gauge indices on the Wigner function (it has two -- 
it taking values in $\mathfrak{g}\otimes\mathfrak{g}^*$, 
\cite{EGV,wletter}). Formally, a solution can be found of the form
\begin{equation}
	W[A,\pi] = e^{{\rm Tr}\int F_{ij}g^{ij} dx}
\end{equation}
where $g^{ij}=g^{ij}(\pi)$ is a quadratic function with values in the
Lie algebra, $g^{ij}_a=d_a^{~bc}\pi_b^i\pi_c^j$ where $d_a^{~bc}$ is
antisymmetric in $bc$ and satisfies
\begin{equation}
	c^c_{~ab}d_c^{~ba} = 8i\hbar^{-2}
\end{equation}
where indices are raised and lowered by the Kronecker delta. Since we already
have one invariant antisymmetric tensor on $\mathfrak{g}$, namely the 
structure coefficient, $c_{~ab}^c$, it is natural to put
\begin{equation}
	d_c^{~ab} = \alpha c_c^{~ab}
\end{equation}
which leads to
\begin{equation}
	\alpha = -\frac{8i}{\hbar^2\kappa}
\end{equation}
where $\kappa=\delta^{ab}\kappa_{ab}$ is the trace of the Cartan-Killing
metric, $\kappa_{ab} = c^c_{~ad}c^d_{~bc}$. For  $\mathfrak{g}=su_n$
for instance, we have $\kappa=-2$, leading to
\begin{displaymath}
	W[A,\pi] = \exp\left(-4i\hbar^{-2}\int c_a^{~bc}F_{ij}^a\pi_b^i\pi_c^j
	dx\right)
\end{displaymath}
For $su_2$, teh Wigner function then involves the dual of the field strength
tensor, whereas in other cases it involves some kind of generalised dual.
In any case, the Wigner function is a kind of ``bi-Gaussian'', i.e., a Gaussian
in either $A$ or $\pi$ with the coeffients depending on the conjugate
variable.\\
It is interesting to note that the present solution cannot be the Wigner
function of a pure state. This is seen as follows. First, the Wigner function
has the form
\begin{displaymath}
	W(q,p) = e^{(aq+bq^2)p^2}
\end{displaymath}
hence the Fourier transform $p\rightarrow y$ looks like
\begin{displaymath}
	W(q,y) = \sqrt{\frac{\pi}{aq+bq^2}}e^{-\frac{1}{4}\frac{y^2}{aq+bq^2}}
\end{displaymath}
By the definition of the Wigner function of a pure state described by a 
wave function $\psi$, we have
\begin{displaymath}
	W(q,0) = |\psi(q)|^2
\end{displaymath}
hence
\begin{displaymath}
	\psi(q) = \left(\frac{\pi}{aq+bq^2}\right)^{1/4} e^{i\theta(q)}
\end{displaymath}
where $\theta$ is purely real. But, on the other hand,
\begin{displaymath}
	W(q,y)=\sqrt{\frac{\pi}{aq+bq^2}}e^{-\frac{1}{4}\frac{y^2}{aq+bq^2}}
	\propto \bar{\psi}(q+\frac{1}{2}y)\psi(q-\frac{1}{2}y)
\end{displaymath}
is only possible if $\theta$ has an imaginary part.\\
The equation for the Wigner function of a Yang-Mills
gauge field, found by Elze, Gyulassy and Vasak, \cite{EGV}, uses second
quantised quantum mechanics and hence operate on the mechanical phasespace
$(q,p)$ and not on the field theoretic one $(A,\pi)$. Consequently, they get
an infinite order partial differential equation whereas we get a finite
order functional differential one. We do also, however, get an infinite order 
equation, since for Yang-Mills theory not all equations of motion are 
constraints, only part of them (corresponding to Gauss' law). The 
remaining equations must be found by the same techniques as used by Elze,
Gyulassy and Vasak, but this time in a purely Hamiltonian framework.\\
One should note that for abelian theories, the second term in (\ref{eq:YMphys})
vanishes and we get the classical requirement ${\cal G}_a\equiv 0$ (since
$W\neq 0$ in general -- it is, a priori, possible to have ${\cal G}_a\neq 0$
at a point $(q,p)$ provided $W(q,p)=0$ of course, hence the zeroes of the 
Wigner function can correspond to points in phasespace at which the classical 
constraints are violated, by continuity of $W$ this can only happen in a 
discrete number of points or in a region disconnected from the rest of
phasespace). Thus, it is the non-abelianness of the gauged algebra, 
$\mathfrak{g}$,
which leads to quantum deformations of the conditions of physical degrees
of freedom.\\
This corresponds to what was proven in \cite{wlie} where it was shown that
a generalised \small{WWM}-formalism exists for a large range of algebraic
structures, the ``deformation'' of the resulting ``classical'' phasespace
(in particular its curvature), being due to the non-commutativity of the
generators of the algebra.\\
It is not surprising that Yang-Mills theory is anomaly free at this level, 
since anomalies tend to appear through the Dirac operator, and hence through
the matter fields, \cite{Nash,Nakahara}. We have only considered Yang-Mills
theory {\em in vacuo} at this stage. The usual anomalies (chiral, parity)
should then reappear only when one computes the Moyal bracket for the fermion
currents. It is already known that the \small{WWM} symbol of an operator
is related to the index of it, and that the Atiyah-Singer index theorem
normally used to express anomalies is intimately related to the entire
\small{WWM}-scheme, \cite{Nest}. One should also take notice of the fact that
the above discussion doesn't take global anomalies into account, such problems
are beyond the scope of the present paper.

\subsection{Gravitation}
For gravity, we will first consider the \small{ADM} constraint algebra, then
the Ashtekar variables and finally make some brief comments on the newly
proposed ``Kucha\v{r} algebra.'', \cite{Kuchar,fotini}\\
The \small{ADM} constraints, \cite{ADM}, are
\begin{eqnarray}
	{\cal H}_\perp(x) &=& g^{-1/2}(\frac{1}{2}\pi^2-\pi^i_j\pi^j_i)+
	\sqrt{g}R = G_{ijkl}\pi^{ij}\pi^{kl}+\sqrt{g}R\\
	{\cal H}_i(x) &=& -2D_j\pi^j_i
\end{eqnarray}
with $g_{ij}$ the 3-metric, $\pi^{ij}$ its conjugate momentum,
\begin{equation}
	\{g_{ij}(x),\pi^{kl}(x')\}_{\rm PB} = \frac{1}{2}(\delta^k_i\delta^l_j
	+\delta^k_j\delta^l_i)\delta(x,x'),
\end{equation}
$R$ the curvature scalar of $g_{ij}$ (i.e., the three dimensional one)
and $g$ the determinant of the 3-metric. The first constraint is known as
the Hamiltonian one, and the last, the ${\cal H}_i$, as the diffeomorphism 
one. The algebra is
\begin{eqnarray}
	\{{\cal H}_\perp(x),{\cal H}_\perp(x')\}_{\rm PB} &=& (g^{ij}(x)
	{\cal H}_j(x)+g^{ij}(x'){\cal H}_j(x'))\delta_{,i}(x,x')\\
	\{{\cal H}_\perp(x),{\cal H}_i(x')\}_{\rm PB} &=& {\cal H}_\perp(x)
	\delta_{,i}(x,x')\\
	\{{\cal H}_i(x),{\cal H}_j(x')\}_{\rm PB} &=& {\cal H}_i(x')\delta_{,j}
	(x,x')+{\cal H}_j(x)\delta_{,i}(x,x')
\end{eqnarray}
where the subscript, $\delta_{,i}$, on the delta function denotes 
partial derivative with respect
to $x^i$. The convention is the standard one in which $\delta(x,x')$ is a
scalar in the first argument and a density in the second (the curved
spacetime Dirac $\delta$ has a $g^{-1/2}$ in it).\\
The algebra of the diffeomorphism constraint will not be deformed as their 
form is ${\cal H}_i \sim \partial\pi + g^2\pi$ and thus has $\omega_3\equiv 
0$. The Hamiltonian constraint, however, has as well a $g\pi^2$ as a $g^2\pi$
term, and will consequently not have vanishing $\omega_3$. We should thus
expect the algebraic relations involving ${\cal H}_\perp$ to receive $\hbar^3$
corrections (but no higher order corrections since no higher powers of $\pi$
are present). This is precisely what we find. Moreover, the Christoffel
symbols and the $\sqrt{g}$ contain, in a Taylor series, the metric to infinite
order, whence we should expect infinite order equations to turn up at some
stage.\\
A straightforward computation yields
\begin{equation}
	[{\cal H}_\perp(x),{\cal H}_\perp(x')]_M = i\hbar\{{\cal H}_\perp(x),
	{\cal H}_\perp(x')\}_{\rm PB}+\hbar^3 k(x,x')
\end{equation}
where
\begin{equation}
	k(x,x') = -\frac{1}{8}i\left((\Xi_{mnab}^{ijkl}\pi^{ab})(x)
	{\cal G}^{mn}_{ijkl}(x') - (x\leftrightarrow x')\right)
\end{equation}
with
\begin{eqnarray}
	\Xi_{mnab}^{ijkl} &\equiv & \frac{\delta^3{\cal H}_\perp}{\delta
	g_{ij}\delta g_{kl}\delta\pi^{mn}}\\
	&=&G_{mnab}(g^{ik}g^{jl}-\frac{1}{2}g^{ij}g^{kl})
	-\frac{1}{2}g^{ij}\left(g_{nb}\delta^k_m\delta^l_a+g_{ma}\delta^k_n
	\delta^l_b-\right.\nonumber\\
	&&\qquad\left.g_{ab}\delta^k_m\delta^l_n-g_{mn}\delta^k_a\delta^l_b
	+g_{bn}\delta^l_m\delta^k_a+g_{am}\delta^l_n\delta^k_b\right)\\
	{\cal G}^{mn}_{ijkl} &\equiv& \frac{\delta^3{\cal H}_\perp}{\delta
	\pi^{ij}\delta\pi^{kl}\delta g_{mn}}\\ 
	&=&-\frac{1}{2}g^{mn}G_{ijkl}+g^{-1/2}\left(g_{jl}(
	\delta^m_i\delta^n_k+\delta^m_k\delta^n_i)+g_{ik}(\delta^m_j\delta^n_l
	+\delta^m_l\delta^n_j)-\right.\nonumber\\
	&&\qquad\left. g_{kl}\delta^m_i\delta^n_j-g_{ij}\delta^m_k\delta^n_l
	\right)
\end{eqnarray}
One should note that $[{\cal H}_\perp,k]_M\neq 0$ hence we get an anomaly which
is not even a central extension of the original algebra. Explicitly
\begin{eqnarray}
	[{\cal H}_\perp,k]_M = i\hbar\{{\cal H}_\perp,k\}_{\rm PB} +i\hbar^3
	\frac{3}{4}\Xi^{ijkl}_{mnab}\pi^{ab}
	\frac{\delta^3 k}{\delta\pi^{ij}\delta g_{kl}\delta
	g_{mn}}\neq 0
\end{eqnarray}
For the \small{ADM} constraints, the structure coefficients depend on the
fields, consequently the anomaly too depends upon $(g,\pi)$.\\
Similarly, the relation mixing ${\cal H}_\perp$ and ${\cal H}_i$ receives a
$\hbar^3$ correction of the form
\begin{displaymath}
	k_i(x,x')\equiv
	-\frac{1}{8}i\frac{\delta^3{\cal H}_\perp(x)}{\delta\pi^{jk}\delta
	\pi^{lm}\delta g_{ab}}\frac{\delta^3{\cal H}_i(x')}{\delta g_{jk}\delta
	g_{lm}\delta\pi^{ab}}
\end{displaymath}
which one easily finds to be 
\begin{equation}
	k_i(x,x') = -\frac{1}{4}i{\cal G}_{jklm}^{pq}\Upsilon_{ipq}^{jklm}
\end{equation}
with
\begin{eqnarray}
	\frac{\delta^3{\cal H}_i(x)}{\delta g_{jk}(x')\delta g_{lm}(x'')
	\delta\pi^{ab}(y)} &=& 2\Upsilon_{iab}^{jklm}(x,x',x'')\delta(x,y)
	\nonumber\\
	&=& \delta^c_{(a}\delta^n_{b)}\delta(x,y)\left\{\delta^l_{(n}
	\delta^m_{i)}\delta(x,x'')\left(-g^{rk}\Gamma^j_{rc}\delta(x,x')
	+\right.\right.\nonumber\\
	&&\left.\frac{1}{2}g^{rs}\left(\delta^j_{(r}\delta^k_{s)}\partial_c
	+\delta^j_{(s}\delta^k_{c)}\partial_r-\delta^j_{(c}\delta^k_{r)}
	\partial_s\right)\delta(x,x')\right)+\nonumber\\
	&&\delta^j_{(n}\delta^k_{i)}\delta(x,x')\left(-g^{rm}\Gamma^l_{rc}
	\delta(x,x'')+\right.\nonumber\\
	&&\left.
	+\frac{1}{2}g^{rs}\left(\delta^l_{(r}\delta^m_{s)}\partial_c
	+\delta^l_{(s}\delta^m_{c)}\partial_r-\delta^l_{(c}\delta^m_{r)}
	\partial_s\right)\delta(x,x'')\right)-\nonumber\\
	&&g_{ni}\delta(x,x'')\left[g^{rl}g^{km}\Gamma^j_{rc}-g^{rk}
	g^{jm}\Gamma^l_{rc}+\right.\nonumber\\
	&&\left.\frac{1}{2}g^{rk}g^{js}\left(\delta^l_{(r}
	\delta^m_{s)}\partial_c+\delta^l_{(s}\delta^m_{c)}\partial_r
	-\delta^l_{(r}\delta^m_{c)}\partial_s\right)-\right.\nonumber\\
	&&\left.\left.\frac{1}{2}g^{rl}g^{sm}\left(\delta^j_{(r}
	\delta^k_{s)}\partial_c+\delta^j_{(s}\delta^k_{c)}\partial_r
	-\delta^j_{(r}\delta^k_{c)}\partial_s\right)\right]\delta(x,x')\right\}
	+\nonumber\\
	&&(c\rightarrow r, n\rightarrow c, r\rightarrow n)
\end{eqnarray}
The spatial diffeomorphism subalgebra spanned by ${\cal H}_i$ does not receive
any quantum corrections since the constraints are only linear in the 
momentum.\\
The set of physical states are defined as the functions $W$ satisfying the 
two infinite order functional differential equations
\begin{eqnarray}
	0 &=& [{\cal H}_\perp,W]_M^+\\
	0 &=& [{\cal H}_i,W]_M^+
\end{eqnarray}
these are infinite order since the Christoffel symbols (and hence the covariant
derivative and the curvature scalar) has an inverse metric in them, similarly
the supermetric $G_{ijkl}$ too has an inverse metric inside. Thus the 
constraints are not polynomial in the metric, but instead ``meromorphic''.\\
Written out more explicitly, the physicality conditions read
\begin{eqnarray}
	0 &=& 2{\cal H}_\perp W +\sum_{k=1}^\infty (-1)^k 2^{-2k-1}\hbar^{2k}
	\left(\frac{\delta^{2k}{\cal H}_\perp}{\delta g_{i_1j_1}...\delta 
	g_{i_{2k-1}j_{2k-1}}\delta g_{mn}}\frac{\delta^{2k}W}{\delta
	\pi^{i_1j_1}...\delta\pi^{i_{2k-1}j_{2k-1}}\delta\pi^{mn}}-
	\right.\nonumber\\
	&&\qquad 2k
	\frac{\delta^{2k}{\cal H}_\perp}{\delta g_{i_1j_1}...\delta 
	g_{i_{2k-1}j_{2k-1}}\delta\pi^{mn}}
	\frac{\delta^{2k}W}{\delta\pi^{i_1j_1}...\delta\pi^{i_{2k-1}j_{2k-1}}
	\delta g_{mn}}+\nonumber\\
	&&\qquad\left.
	k(2k-1)\frac{\delta^{2k}{\cal H}_\perp}{\delta g_{i_1j_1}
	...\delta g_{i_{2k-2}j_{2k-2}}\delta\pi^{i_{2k-1}j_{2k-1}}
	\delta\pi^{mn}}\frac{\delta^{2k}W}{\delta\pi^{i_1j_1}...
	\delta\pi^{i_{2k-2}j_{2k-2}}\delta g_{i_{2k-1}j_{2k-1}}\delta
	g_{mn}}\right)\nonumber\\
	&&\\
	0 &=& 2{\cal H}_i W + \sum_{k=1}^\infty (-1)^k 2^{-2k-1}\hbar^{2k}
	\left(\frac{\delta^{2k}{\cal H}_i}{\delta g_{i_1j_1}...\delta 
	g_{i_{2k}j_{2k}}}\frac{\delta^{2k} W}{\delta\pi^{i_1
	j_1}...\delta\pi^{i_{2k}j_{2k}}}-\right.\nonumber\\
	&&\qquad\left.
	2k\frac{\delta^{2k}{\cal H}_i}{\delta g_{i_1j_1}...\delta 
	g_{i_{2k-1}j_{2k-1}}\delta\pi^{i_{2k}j_{2k}}}
	\frac{\delta^{2k}W}{\delta\pi^{i_1j_1}...\delta\pi^{i_{2k-1}j_{2k-1}}
	\delta g_{i_{2k}j_{2k}}}\right)
\end{eqnarray}
Since the constraints for gravity in the \small{ADM} formalism are 
non-polynomial the equations defining the physical state space become infinite
order. If one assumes the Wigner function to be analytic in $\hbar$, one
can Taylor expand it $W=\sum_{n=0}^\infty \hbar^n W_n$, and arrive at the
following recursive formulas for the $n$'th order coefficients, $W_n$
\begin{eqnarray}
	0 &=& {\cal H}_\perp W_0 = {\cal H}_i W_0\\
	0 &=& 2{\cal H}_\perp W_N+\sum_{k=1}^{[N/2]} (-1)^k2^{-2k-1}
	\left(\frac{\delta^{2k}{\cal H}_\perp}{\delta g_{i_1j_1}...
	\delta g_{i_{2k-1}j_{2k-1}}\delta\pi^{mn}}
	\frac{\delta^{2k}}{\delta\pi^{i_1j_1}...\delta\pi^{i_{2k-1}j_{2k-1}}
	\delta g_{mn}}-\right.\nonumber\\
	&&\qquad\left. 2k\frac{\delta^{2k}{\cal H}_\perp}{\delta g_{i_1j_1}
	...\delta g_{i_{2k-2}j_{2k-2}}\delta\pi^{i_{2k-1}j_{2k-1}}
	\delta\pi^{mn}}\frac{\delta^{2k}}{\delta\pi^{i_1j_1}...
	\delta\pi^{i_{2k-2}j_{2k-2}}\delta g_{i_{2k-1}j_{2k-1}}\delta
	g_{mn}}\right) W_{N-2k}\nonumber\\
	&&\\
	0 &=& 2{\cal H}_iW_N+\sum_{k=1}^{[N/2]}(-1)^k 2^{-2k-1}
	\frac{\delta^{2k}{\cal H}_i}{\delta g_{i_1j_1}...\delta 
	g_{i_{2k-1}j_{2k-1}}\delta\pi^{mn}}
	\frac{\delta^{2k}W_{N-2k}}{\delta\pi^{i_1j_1}...
	\delta\pi^{i_{2k-1}j_{2k-1}}\delta g_{mn}}
\end{eqnarray}
where $k,N\geq 1$. These equations are not enough to completely specify the
Wigner functions.\\ 
The conclusion so far is then that in the \small{ADM} formalism gravity is
anomalous when one attempts a deformation quantisation. The question is, then,
whether one can lift these anomalies or not.\\
Letting ${\cal H}_0={\cal H}_\perp$ and $\mu=(0,i)$ we then want quantum
constraints $H_\mu$ satisfying
\begin{equation}	
	[H_\mu(x),H_\nu(x')]_M = c_{\mu\nu}^\rho(x,x')H_\rho(x)
	+d_{\mu\nu}^\rho(x,x')H_\rho(x')
\end{equation}
where the structure coeffcients $c_{\mu\nu}^\rho,d_{\mu\nu}^\rho$ depend
on the coordinates only through Dirac delta functions (and their derivatives)
and directly through the phasespace variables. Because of the complicated
nature of the constraints in the \small{ADM} formalism I have not been
able to find a good set of quantum constraints.\\
We saw that the anomalous nature of the quantum deformed algebra of the
contraints in the \small{ADM} formalism were due to the constraints being
non-polynomial. It is therefore interesting ot consider another formulation,
the Ashtekar variables \cite{Ashtekar}, where the constraints {\em are}
polynomials. In this formulation the canonical coordinates are a complex
$su_2$-connection $A_i^a$ and its momentum (a densitised {\em dreibein}) 
$E_a^i$, and the constraints are
\begin{eqnarray}
	{\cal H} &=& F_{ij}^a E^i_bE^j_c\varepsilon^{bc}_{~~~a}\\
	{\cal G}_a &=& D_iE^i_a\\
	{\cal D}_i &=& F_{ij}^aE^j_a
\end{eqnarray}
It will turn out that in these variables the anomaly is much simpler, namely
merely a central extension.\\
Since the Ashtekar variables bring out the analogy between general relativity 
and (complexified) Yang-Mills theory due to the isomorphism $so(3,1)\simeq
su_2\otimes\mathbb{C}$, we can use our knowledge of the deformation 
quantisation of
Yang-Mills systems to see that only the following two brackets can receive
any quantum corrections, and these only to lowest order
\begin{eqnarray}
	\left[{\cal H}(x),{\cal H}(x')\right]_M 
	&=& i\hbar\{{\cal H}(x),{\cal H}(x')
	\}_{\rm PB}+\frac{3}{4}i\hbar^3\left(\frac{\delta {\cal H}(x)}
	{\delta A^2\delta E}\frac{\delta{\cal H}(x')}{\delta E^2\delta A}
	-(x\leftrightarrow x')\right)\nonumber\\
	&&\\
	\left[{\cal H}(x),{\cal D}_i(x')\right]_M 
	&=& i\hbar\{{\cal H}(x),{\cal D}_i(x')
	\}_{\rm PB} - \frac{3}{4}i\hbar^3\frac{\delta^3{\cal H}(x)}{\delta E^2
	\delta A}\frac{\delta^3{\cal D}_i(x')}{\delta A^2 \delta E}	
\end{eqnarray}
where we have suppressed the indices on the $A,E$. An explicit and 
straightforward computation gives
\begin{eqnarray}
	\omega_3({\cal H}(x),{\cal H}(x')) &=& -12 i\delta(x,x')\left(E^j_a(x)
	A_j^a(x)-E^j_a(x')A_j^a(x')\right)\\
	\omega_3({\cal H}(x),{\cal D}_m(x')) &=& 9i\delta_{,m}(x,x')
\end{eqnarray}
We notice that the first of these vanish in the sense of distributions, hence
the only quantum correction is the constant (w.r.t. the phasespace variables)
$9i\delta_{,m}(x,x')$. Consequently, the anomalous nature of gravity shows
itself in the Ashtekar variables simply in a central extension of the
constraint algebra (similar to a Schwinger term in current algebra).
\begin{equation}
	[{\cal H}(x),{\cal D}_i(x') ]_M = i\hbar\{{\cal H}(x),{\cal D}_i(x')
	\}_{\rm PB} -9i\hbar^3\delta_i(x,x')
\end{equation}
As we have seen earlier such central extensions can be
``lifted'' by means of a redefinition of the quantum constraints involving
negative powers of $\hbar$.\\
Furthermore, since the constraints are polynomial
in the phasespace variables the equations defining the physical state
space, $\tilde{\cal C}_{\rm phys}$, become finite order differential
equations. Explicitly, since the constraints are at most quartic in the
phasespace variables
\begin{eqnarray}
	0=[{\cal H},W]_M^+ &=& 2{\cal H}W-\frac{1}{2}\hbar^2\left(E_b^kE_v^l
	\epsilon_a^{~bc}\epsilon^a_{~ef}\frac{\delta^2W}{\delta E^k_e\delta
	E^l_f}-\right.\nonumber\\
	&&2\epsilon_a^{~bc}\left(-\delta^a_e(\delta^k_i\partial_j-\delta^k_j
	\partial_i)+\epsilon^a_{~pq}(\delta^p_e\delta^k_iA^q_j
	+\delta^q_e\delta^k_jA_i^p)\right)(\delta^i_l\delta^b_fE^j_c
	+\delta^j_l\delta^c_fE^i_b)\frac{\delta^2W}{\delta E_e^k\delta A^f_l}
	+\nonumber\\
	&&\left.\epsilon_a^{~bc}F_{ij}^a\frac{\delta^2W}{\delta A_i^b
	\delta A_j^c}\right)+\frac{5}{4}\hbar^4\epsilon_{bc}^{~~a}
	\epsilon_a^{~ef}\frac{\delta^4W}{\delta E^k_b\delta E^l_c\delta A^e_k
	\delta A^f_l}\\
	0=[{\cal D}_i,W]_M^+ &=&2{\cal D}_iW-\frac{1}{2}\hbar^2\left(
	\epsilon^a_{~ef}E^j_a\frac{\delta^2W}{\delta E_e^i\delta E^j_f}-
	\right.\nonumber\\
	&&2\left(-\delta^a_e(\delta^k_i\partial_j-\delta^k_j\delta_i)
	+\epsilon^a_{~mn}(\delta^m_e\delta^k_iA^m_j+\delta^n_e\delta^k_j
	A^m_i)\right)\frac{\delta^2W}{\delta E^k_e\delta A^a_j}\\
	0=[{\cal G}_a,W]_M^+ &=& 2{\cal G}_aW + \frac{1}{4}i\hbar^2\delta^j_k
	\epsilon^c_{~ab}\frac{\delta^2W}{\delta A^c_j\delta A^k_b}
\end{eqnarray}
These coupled equations constitute the equations for the Wigner function
for Ashtekar gravity in vacuum.\\
So far the Ashtekar variables have turned out to be much more tractable than
the \small{ADM}-approach; the ``anomaly'' was simply a Schwinger term and the
equations for the physical states are finite order. There is one problem, 
however, which one seldom take into account
in this formalism, namely the reality conditions. The physical variables
have to correspond to a real metric, not a complex one. This condition is a
second class constraint, \cite{AshReal}, and this is where the real problems
with the Ashtekar formalism lies. But since we have seen that second class
constraints can be treated rather easily in this formalism, the conclusion must
be that the Ashtekar approach to gravity, all things considered, is the most
fruitful one.\\
As a final comment, let us digress on the recently proposed ``Kucha\v{r} 
algebra'', \cite{Kuchar}. Here the Hamiltonian constraint ${\cal H}_\perp$
of the \small{ADM} approach is replaced by an abelian constraint $\cal K$ (a
scalar density of weight $\omega$),
leading to the algebra of $C^\infty(\Sigma)\ltimes{\rm Diff}(\Sigma)$, i.e.,
\begin{eqnarray}
	\{{\cal K}(x),{\cal K}(x')\}_{\rm PB} &=& 0\\
	\{{\cal K}(x),{\cal H}_i(x')\}_{\rm PB} &=& {\cal K}_{,i}\delta(x,x')
	+\omega {\cal K}\delta_{,i}(x,x')\\
	\{{\cal H}_i(x),{\cal H}_j(x')\}_{\rm PB} &=& {\cal H}_i(x)\delta_{,j}
	(x,x')+{\cal H}_j(x')\delta_{,i}(x,x')
\end{eqnarray}
There is an appealing interpretation of this algebra in terms of fibrebundles
over a three-manifold, \cite{fotini}, but since the only known representations
of this algebra are the ones formed from the \small{ADM} constraints, such as
\begin{displaymath}
	{\cal K} = {\cal H}_\perp^2-{\cal H}_i{\cal H}_j g^{ij}\\
\end{displaymath} 
and even more non-linear expressions (the general solution being found in
\cite{Kuchar}), we will not be able to compute the Moyal brackets in any
satisfactory way -- they will {\em a priori} involve infinite order
differentials. There is, however, still the hope that some of the solutions,
i.e., some of the explicit expressions for the ${\cal K}$'s, will turn out
to simplify the Moyal bracket and will hence be somehow favoured. At the
present, it has to be admitted, though, that this is a rather faint hope.\\
It is not enough for a constraint algebra to be nice, it is also important
for the representation of the constraint algebra as functions on phasespace
to be sufficiently simple, in order for the deformation quantisation scheme
to be really tractable. This of course holds in practice for any quantisation
scheme, but contrary to some other schemes, deformation quantisation is at
least in principle able to handle constraints of arbitrary complecity -- it
is merely a matter of computational convenience (or laziness).\\
In a sequel paper we will concentrate on the deformation quantisation of
gravity, and we will interpret the set of quantum constraints in the
Ashtekar variables, find the relationship to the loop formalism and knot
invariants and finally find an explicit solution related to topological
field theory, \cite{defgrav}.

\section{Underlying Geometrical Structure}
In the present picture one quantizes a classical theory, as described by an
algebra of observables ${\cal A}_0 = C^\infty(\Gamma)$, by deforming it
to obtain a noncommutative algebra ${\cal A}_\hbar\simeq {\cal A}_0\otimes
\mathbb{C}((\hbar))$ 
(isomorphism as vectorspaces, not as algebras), with a new,
twisted, product $f,g\mapsto f*g=fg+O(\hbar)$.\\
We can consider ${\cal A}_\hbar$ as a ``field'' (in the language of 
$C^*$-algebras) over the real axis, parametrised by $\hbar\in\mathbb{R}$. 
This is the
point of view taken by Nest et al, \cite{Nest}. Alternatively, we can consider
a {\em sheaf}, \cite{Warner, alggeom}, ${\cal A}$ over the topological
space $X=\mathbb{R}$, where the stalks are given by
\begin{equation}
	{\cal A}_x = \left\{\begin{array}{ll}
		0 & x<0\\
		{\cal A}_0 & x=0\\
		{\cal A}_\hbar & x>0
	\end{array}\right.
\end{equation}
where $\hbar$ is taken to be a free parameter, $\hbar:=x$. This is obviously a
fine sheaf, since a partition of the identity trivially exists, furthermore,
it is a sheaf which is almost constant. This expresses quantisation as 
a sheafication process.\\
We have already commented a bit on some underlying cohomological structure,
here we will elaborate on this. First, each $f\in{\cal A}_\hbar$ can be
written as a Laurent series (we can without loss of generality assume only one
negative power of $\hbar$ to be present)
\begin{equation}
	f=f_{-1}\hbar^{-1}+f_0+f_1\hbar+f_2\hbar^2+...\qquad f_i\in{\cal A}_0
\end{equation}
This leads to the definition of a series of natural differentials, each with
their own interpretation
\begin{eqnarray*}
	\delta_0 &:& f\rightarrow f_0\\
	\delta_q &:& f\rightarrow f_1\hbar+\hbar^2 f_2+...\\
	\delta_- &:& f\rightarrow f_{-1}\\
	\delta_+ &:& f\rightarrow f_1
\end{eqnarray*}
where $\delta_\pm$ are inspired by the corresponding definition for the
Weil complex, \cite{Nest}. Clearly $\delta_0^2=\delta_+^2=\delta_-^2=0$ so
these three all define cohomology theories by considering the trivial
complex
\begin{equation}
	0\rightarrow {\cal A}_\hbar \stackrel{\delta}{\rightarrow}
	{\cal A}_\hbar \stackrel{\delta}{\rightarrow}
	{\cal A}_\hbar \rightarrow ...
\end{equation}
where $\delta$ denotes any of the differentials $\delta_\pm,\delta_0$. The
cohomologies are trivial
\begin{equation}
	H^0(\delta_0) = {\cal A}_0\qquad H^0(\delta_\pm) = \delta_\pm{\cal 
	A}_\hbar
\end{equation}
thus, $H^0(\delta_0)$ gives the classical algebra back, whereas $H^0(\delta_-)$
gives the anomalous part.\\
A more interesting complex is
\begin{equation}
	0\rightarrow {\cal A}_\hbar \stackrel{\delta_0}{\rightarrow}
	{\cal A}_\hbar \stackrel{\delta_q}{\rightarrow}
	{\cal A}_\hbar \rightarrow ...
\end{equation}
with alternating $\delta_0,\delta_q$. This is indeed a complex since obviously
$\delta_0\delta_q+\delta_q\delta_0=0$. This has the same cohomology as the
$\delta_0$-complex, since ${\rm Im}\delta_0={\rm Ker}\delta_q$.\\
But, instead of considering the cohomology of a single stalk, it is better to
consider the cohomology of the entire sheaf. This will also allow us to take
the physicality condition into account. This condition can be rewritten as
\begin{equation}
	sf:=\left[\eta^\alpha\Phi_\alpha,f\right]_M=0
\end{equation}
where now $\eta^\alpha$ a constant Grassmann numbers and the Moyal bracket is
graded appropriatly - when $f$ is Grassmann even this becomes the anti-Moyal
bracket. Then, $s^2=0$, and $s$ is very similar to the BRST differential.
By extending the algebras ${\cal A}_x$ to include the constant Grassmann
parameters $\eta^\alpha$, we get a complex of sheaves
\begin{equation}
	s: {\cal A}^q\rightarrow {\cal A}^{q+1}
\end{equation}
where ${\cal A}^q$ denotes the sheaf obtained from $\cal A$ by tensoring
\begin{equation}
	{\cal A}^q = {\cal A}\otimes \underbrace{\mathbb{G}\otimes\mathbb{G}
	\otimes...\otimes \mathbb{G}}_q
\end{equation}
where $\mathbb{G}$ denotes the space of the $\eta_\alpha$-parameters. This 
leads us naturally to the concept of {\em hypercohomology}, \cite{alggeom}. 
Let $\{I\}$ be a covering of $X=\mathbb{R}$ by open sets, i.e., $\{I\}$ 
consists of open interval $]a,b[$ such that their union gives the entire 
real axis. Given such a covering we can define \v{C}ech cochains with values
in the sheaf complex ${\cal A}^*$, the set of these is denoted by $\check{C}^p
(I,{\cal A}^*)$ and is defined by, \cite{Warner},
\begin{equation}
	\check{C}^p(I,{\cal A}^*) = \{f: I^{p+1}\rightarrow {\cal A}^*\}
\end{equation}
The \v{C}ech cochains come with a differential, $\check{\delta}$, given by
\begin{equation}
	\check{\delta}f(I_0,...,I_{p+1}) = f(I_0\cap I_{p+1},I_1,...,I_p)
	+\sum_{i>j} (-1)^{i+j}f(I_0,...,\hat{I}_i,...,\hat{I}_j,... I_i\cap 
	I_j)
\end{equation}
We hence have a bicomplex,  $\check{C}^p(I,{\cal A}^q)$, the cohomology of
the total complex is denoted by simply $H^*(I,{\cal A}^*)$. 
Now, any refinement of the cover
induces homomorphisms among the cochains and hence also among the cohomology
modules, consequently, we can perform the limit of ever finer coverings to 
obtain the hypercohomology
\begin{equation}
	\mathbb{H}^*(X,{\cal A}^*) := \lim_I H^*(I,{\cal A}^*)
\end{equation}
Since the sheaf is fine, we have
\begin{equation}
	\mathbb{H}^p(X,{\cal A}^q) = 0\qquad p\neq 0
\end{equation}
On the other hand
\begin{equation}
	\mathbb{H}^0(X,{\cal A}^q) = \lim_I {\rm Ker}(\check{\delta}+s)
\end{equation}
In ghost number zero, we therefore have
\begin{equation}
	\mathbb{H}^0(X,{\cal A}^0) = {\cal C}_{\rm phys}
\end{equation}
This method is compatible with BRST-techniques. One simply defines the 
quantisation by means of the abovementioned sheafication procedure. This 
remedies the usual shortcoming of canonical quantisation techniques, such as
the BRST, by avoiding finding an operator realisation. One just considers
the BRST symmetry as a classical symmetry, and then deforms it by replacing the
BRST-complex by its sheafication over the real axis.

\section{Conclusion}
We have studied the deformation (i.e., Moyal) quantisation of Hamiltonian
systems with constraints. Instead of looking for an operator correspondence
$q_i,p^i\rightarrow \hat{q}_i,\hat{p}^i$ with the usual rule
$\{f,g\}_{\rm PB}\rightarrow \frac{1}{i\hbar}[\hat{f},\hat{g}]$, we keep the
classical phasespace but replace the Poisson bracket with the Moyal bracket,
$[f,g]_M = i\hbar\{f,g\}_{\rm PB}+O(\hbar^2)$. It is know that this is always
possible. By introducing sufficiently many phasespace variables and 
constraints one can assume the classical phasespace to be flat (i.e., to have 
a global patch of Darboux coordinates, $\{q_i,p^j\}_{\rm PB} = \delta_i^j$).\\
A priori one have to replace the classical constraints $\phi_a$ by quantum
deformed versions $\Phi_a = \sum_n\hbar^n\Phi_a^{(n)}$ with $\Phi_a^{(0)} =
\phi_a$ in order to keep the constraint algebra when one replaces Poisson
brackets by Moyal ones. We saw that for Yang-Mills systems we could take
$\Phi_a=\phi_a$, but for systems where the constraints are more than cubic
in the phasespace variables, one would in general have $\Phi_a\neq \phi_a$.\\
Second class constraints could be handled by a similar device, turning them 
into a an Abelian algebra under Moyal brackets. This was illustrated by the
case of the $O(N)$ non-linear $\sigma$-model.\\
The dynamics of a constrained Hamiltonian system was captured by the Wigner
function, where the classical condition $\phi_a=0$ was replaced not by the
operator equation $\hat{\phi}_a|\Psi\rangle=0$ but by the algebraic condition
$[\Phi_a,W]_M^+=0$ resembling the \small{BRST} approach.\\
Various problems can arrise: (1) the appearance of second class constraints,
(2) the appearance of first class constraints not in involution with the
Hamiltonian, (3) the appearance of anomalies. It has been shown in this paper
how all these problems can be remedied in deformation quantisation, provided
one is willing to include negative powers of the deformation paramater $\hbar$
in the expansions of the quantum objects and to let the structure coefficients
receive $\hbar$-corrections. Hence, upon replacing $
{\cal A}_\hbar= {\cal A}\otimes\mathbb{C}[[\hbar]]$ 
($\cal A$-valued formal power
series in $\hbar$) by ${\cal A}_\hbar = {\cal A}\otimes\mathbb{C}((\hbar))$ 
(the
set of $\cal A$-valued formal Laurent series in $\hbar$). This of course 
implies that the naive classical limit $\hbar\rightarrow 0$ becomes singular.
The correct classical limit appears only upon taking principal values. Thus,
the naive classical limit $A_{\rm clas} = \lim_{\hbar\rightarrow 0} A$ is
replaced by
\begin{equation}
	A_{\rm clas} = PP\lim_{\hbar\rightarrow 0}A = \delta_0A
\end{equation}
where $PP$ denotes principal part and where $\delta_0$ is the differential
defined in the previous section.\\
With this definition the classical part is {\em still} the $\hbar^0$ 
component, there is just a subtle difference between picking out this
component and taking the limit $\hbar\rightarrow 0$.\\
With this {\em caveat}, deformation quantisation emerges as a very powerful
quantisation scheme indeed. In particular, on should note the very explicit
form the relations determining the $\hbar$-modification to the classical
objects take -- an explicit form allowing us in certain examples to arrive
at an explicit solution to all orders in $\hbar$. Consequently, deformation
quantisation is a very constructive approach to quantisation. The {\em caveat}
about the principal part also illuminates the nature of the classical limit,
and in particular how ill-behaved {\em classical} theories can arrise from 
well-behaved {\em quantum} ones due to the singularity of the limit. 
Alternatively,
one can view this as illustrating the quantum ``smearing out'' of problems
in the classical theory. With this in mind, it would be very interesting to
study the application of deformation quantisation to quantum chaos.\\
Yang-Mills theory was then quantised in this scheme and we saw that in the 
absence of matter, the theory was anomaly free.\\
Gravitation was treated in three different manners, first the standard 
\small{ADM} approach, secondly in the Ashtekar variables and finally with a
recently proposed new constraint algebra. We found that gravity in the
\small{ADM} formalism was anomalous and lead to infinite order equations
for the physical states -- both problems stemming from the non-polynomial
nature of the constraints. The Ashtekar variables, however, turned out to
aquire merely a central extension, which can be lifted. Moreover, since
these constraints were polynomial, the equations picking out physical
states became finite order.


\begin{thebibliography}{10}
\bibitem{geomq} N. Woodhouse, {\em Geometric quantisation}, (Oxford University
Press, Oxford, 1980).
\bibitem{deform} F. Bayen, M. Flato, C. Fronsdal, A. Lichnerowicz, D.
Sternheimer: Ann. Phys. {\bf 110} (1978) 61; Ann. Phys. {\bf 110} (1978) 111;
M. DeWilde, P. B. A Lecomte, Lett. Math. Phys. {\bf 7} (1983) 487; B. Fedosov,
J. Diff. Geom. {\bf 40} (1994) 213.
\bibitem{Tzan} C. Tzanakis, A. Dimakis: q-alg/9605018
\bibitem{wigner} R. L. Liboff, {\em Kinetic Theory, Classical, Quantum and 
Relativistic Descriptions} (Prentice Hall, Englewood Cliffs, NJ, 1990); W. A. 
van Leeuwen, Ch. G. van Weert and S. R. de Groot, {\em Relativistc Kinetic 
Theory} (North-Holland, Amsterdam, 1980)
\bibitem{Jacobson} N. Jacobson, {\em Lie Algebras}, (Dover, New York, 1962).
\bibitem{EGV} H.-Th. Elze, M. Gyulassy and D. Vasak, Nucl. Phys. {\bf B276}
706 (1986); Phys. Lett. B{\bf 177} 402 (1986); D. Vasak, M. Gyulassy and
H.-Th. Elze, Ann. Phys. (NY) {\bf 173}, 462 (1987).
\bibitem{wletter} F. Antonsen, Phys.Rev.D{\bf 56} (1997) 920.
\bibitem{wlie} F. Antonsen, Int. J. Theor. Phys. {\bf 37} (1998) 697, 
quant-ph/96098142; short version in {\em Proceedings of the
Third International Wigner Symposium, Guadalajara 1995}, N. M. Atakishiyev,
Th. Seligmann and K. B. Wolf (eds) (World Scientific, Singapore, 1996)
\bibitem{qft} P. Ramond, {\em Field Theory: A Modern Primer/2ed}, 
(Addison-Wesley, Redwood City, CA, 1989); C. Itzykson, J.-B. Zuber, 
{\em Quantum Field Theory} (McGraw-Hill, New York 1980).
\bibitem{Dahl} A. Grossmann, Commun. Math. Phys. {\bf 48}, 191 (1976); A. 
Royer, Phys. Rev. A{\bf 15}, 449 (1977); J.-P. Dahl, Phys. Scr. {\bf 25}, 499
(1982).
\bibitem{ADM} R. Arnowitt, S. Deser, C. Misner in {\em Gravitation: An
Introduction to Current Research}, L. Witten (ed) (John Wiley, New York 1962);
good reviews include K. Kucha\v{r} in {\em Relativistic Astrophysics and
Cosmology}, W. Israel (ed) (Reidel, Dordrecht 1973); C. J. Isham in
{\em Quantum Gravity} C.J. Isham, R. Penrose, D.W. Sciama (eds) (Clarendon,
Ocford 1975); papers by J. Isenberg and K. Nestor, C. Teitelboim and P. G.
Bergmann \& A. Komar in {\em General Relativity and Gravitation}, A. Held
(ed) (Plenum, New York 1980). 
\bibitem{Nash} C. Nash, {\em Differential Topology and Quantum Field Theory},
(Academic, London, 1991).
\bibitem{Nakahara} M. Nakahara, {\em Geometry, Topology and Physics} (IOP,
Bristol, 1990).
\bibitem{Kuchar} K. V. Kucha\v{r}, Phys. Rev. D{\bf 43} (1991) 3332; Phys. Rev.
D{\bf 44} (1991) 43; K. V. Kucha\v{r}, J. D. Romano, Phys. Rev. D{\bf 51} 
(1995) 5579; F. G. Markopoulou, Class. Quant. Grav. {\bf 13} (1996) 2577.
\bibitem{fotini} F. Antonsen, F. G. Markopoulou, gr-qc/9702046.
\bibitem{Ashtekar} A. Ashtekar, Phys. Rev. Lett. {\bf 57}, 2244 (1986); Phys.
Rev. D{\bf 36}, 1587 (1987).
\bibitem{AshReal} H. A. Morales-Tecotl, L. F. Urratia, J. D. Vergara, Class.
Quant. Grav. {\bf 13} (1996) 2933.
\bibitem{brst} M. Henneaux, Phys. Rep. {\bf 126} (1986) 1.
\bibitem{SD} H. Kleinert, S.V. Shabanov, Phys. Lett. {\bf A232} (1997) 327.
\bibitem{Hamachi} K. Hamachi, Lett. Math. Phys. {\bf 40} (1997) 257.
\bibitem{other} G. Jorjadze, J. Math. Phys. {\bf 38} (1997) 2851; G. Junker,
J.R. Klauder, quant-ph/9708027; J.R. Klauder, quant-ph/9612025.
\bibitem{Plebanski} H. Garcia-Compe\'{a}n, J. F. Pleba\'{n}ski, N. 
Quiroz-P\'{e}rez, hep-th/9610248.
\bibitem{selfdual} I. A. B. Strachan, Phys. Lett. {\bf B282} (1992) 63;
K. Takasaki, J. Geom. Phys. {\bf 14} (1994) 111
\bibitem{Nest} R. Nest, B. Tsygan, Commun. Math. Phys. {\bf 172} (1995) 223;
Adv. in Math. {\bf 113} (1995) 151; G. A. Elliot, T. Natsume, R. Nest, 
K-Theory {\bf 7} (1993) 409.
\bibitem{DerVer} T. Dereli, A. Ver\c{c}in, quant-ph/9707040.
\bibitem{Gozzi} E. Gozzi, M. Reuter, Int. J. Mod. Phys. A{\bf 9}, no 32. 
(1994) 5801; Int. J. Mod. Phys. A{\bf 9}, no. 13 (1994) 2191; E. Gozzi,
Nucl. Phys. B (Proc. Suppl.) {\bf 57} (1997) 223.
\bibitem{Kontsevich} M. Kontsevich, q-alg/9709040.
\bibitem{defgrav} F. Antonsen, ``Deformation Quantisation of Gravity'',
gr-qc/9712012.
\bibitem{Warner} F. K. Warner, {\em Foundations of Differentiable Manifolds
and Lie Groups}, Springer-Verlag, New York, 1983.
\bibitem{alggeom} P. Griffiths, J. Harris, {\em Principles of Algebraic
Geometry}, Wiley, New York, 1978.
\end{thebibliography}
\end{document}